# Impact Erosion Model for Gravity-Dominated Planetesimals

Hidenori Genda[1*], Tomoaki Fujita[2], Hiroshi Kobayashi[3],
Hidekazu Tanaka[4], Ryo Suetsugu[5], Yutaka Abe[2]

[1]**Earth-Life Science Institute, Tokyo Institute of Technology,
Ookayama, Meguro-ku, Tokyo, 152-8550, Japan**
[2]**Department of Earth and Planetary Science, The University of Tokyo,
Hongo, Bunkyo-ku, Tokyo, 113-0033, Japan**
[3]**Department of Physics, Graduate School of Science, Nagoya University,
Furo-cho, Chikusa-ku, Nagoya, 464-8602, Japan**
[4]**Astronomical Institute, Tohoku University,
Aramaki, Aoba-ku, Sendai 980-8578, Japan**
[5]**School of Medicine, University of Occupational and Environmental Health,
Iseigaoka, Yahata, Kitakyusyu, 807-8555, Japan**

*Corresponding author: Hidenori Genda

Mailing address: Earth-Life Science Institute, Tokyo Institute of Technology,

2-12-1-IE-14 Ookayama, Meguro-ku, Tokyo, 152-8550, Japan

Email: genda@elsi.jp, Phone: +81-3-5734-2887

**Abstract** Disruptive collisions have been regarded as an important process for planet formation, while non-disruptive, small-scale collisions (hereafter called erosive collisions) have been underestimated or neglected by many studies. However, recent studies have suggested that erosive collisions are also important to the growth of planets, because they are much more frequent than disruptive collisions. Although the thresholds of the specific impact energy for disruptive collisions ($Q_{RD}{}^*$) have been investigated well, there is no reliable model for erosive collisions. In this study, we systematically carried out impact simulations of gravity-dominated planetesimals for a wide range of specific impact energy ($Q_R$) from disruptive collisions ($Q_R \sim Q_{RD}{}^*$) to erosive ones ($Q_R \ll Q_{RD}{}^*$) using the smoothed particle hydrodynamics method. We found that the ejected mass normalized by the total mass ($M_{ej}/M_{tot}$) depends on the numerical resolution, the target radius ($R_{tar}$) and the impact velocity ($v_{imp}$), as well as on $Q_R$, but that it can be nicely scaled by $Q_{RD}{}^*$ for the parameter ranges investigated ($R_{tar}$ = 30–300 km, $v_{imp}$ = 2–5 km/s). This means that $M_{ej}/M_{tot}$ depends only on $Q_R/Q_{RD}{}^*$ in these parameter ranges. We confirmed that the collision outcomes for much less erosive collisions ($Q_R < 0.01\ Q_{RD}{}^*$) converge to the results of an impact onto a planar target for various impact angles ($\theta$) and that $M_{ej}/M_{tot} \propto Q_R/Q_{RD}{}^*$ holds. For disruptive collisions ($Q_R \sim Q_{RD}{}^*$), the curvature of the target has a significant effect on $M_{ej}/M_{tot}$. We also examined the angle-averaged value of $M_{ej}/M_{tot}$ and found that the numerically obtained relation between angle-averaged $M_{ej}/M_{tot}$ and $Q_R/Q_{RD}{}^*$ is very similar to the cases for $\theta = 45°$ impacts. We proposed a new erosion model based on our numerical simulations for future research on planet formation with collisional erosion.

# 1. Introduction

Collisions are one of the most fundamental processes in planet formation. In each stage of planet formation, many collisions take place continuously, and the impact scale varies from micrometers to the 1000-km scale. For example, accumulation of micrometer dust occurs early in the stage of planetesimal formation (e.g., Weidenschilling, 1980, Wada et al., 2009), the accretion of km-sized planetesimals occurs in the stage of protoplanet formation (e.g., Wetherill and Stewart, 1989; Kokubo and Ida, 1996), and giant impacts of 1000-km protoplanets occur in the last stage of terrestrial planet formation (e.g., Chambers and Wetherill, 1998, Kokubo and Genda, 2010).

If colliding bodies simply merge, collisions promote planet growth. However, collision phenomena are not so simple. For example, in the stage of planetesimal formation, collisions between dust aggregates accelerated by turbulence in a protoplanetary disk can be so destructive that the dust aggregates break into fragments instead of growing (Weidenschilling, 1984; Wada et al., 2008). The stage of protoplanet formation involves a similar problem. Protoplanets become massive, their stirring increases the random velocity of surrounding planetesimals, and collisions between planetesimals become more destructive. As meter-sized fragments resulting from the destructive collisions between planetesimals are removed by rapid radial drift due to gas drag in the protoplanetary disk, the depletion of bodies accreting onto protoplanets can stall protoplanet growth (Inaba et al., 2003; Kenyon and Bromley, 2008; Kobayashi et al., 2010, 2011). Conversely, the radial drift of fragments resulting from destructive collisions can also accelerate protoplanet growth at a pressure maximum in the protoplanetary disk (e.g., Kobayashi et al., 2012; Zhu et al., 2012).

Depending on the impact energy of two colliding objects, the collision outcomes can be divided roughly into two categories: disruptive collision and non-disruptive collision. Disruptive collision, which occurs in high-energy impacts, extensively destroys the colliding bodies. In contrast, non-disruptive collision produces a mass of ejecta that is much smaller than the total mass of the colliding bodies. In the literature on planetary collisions, the specific energy of impact is often used to discuss collisional outcomes. Here, we use the following expression of the specific impact energy from Leinhardt and Stewart (2012):

$$Q_\text{R} = \left(\frac{1}{2}M_\text{tar}V_\text{tar}^2 + \frac{1}{2}M_\text{imp}V_\text{imp}^2\right)\bigg/M_\text{tot} = \left(\frac{1}{2}\mu v_\text{imp}^2\right)\bigg/M_\text{tot} \quad (1)$$

where $M_\text{tar}$ and $M_\text{imp}$ are the mass of the target and impactor (here $M_\text{tar} > M_\text{imp}$, and $M_\text{tot} = M_\text{imp} + M_\text{tar}$), respectively, $V_\text{tar}$ and $V_\text{imp}$ are the velocity of the target and impactor in the frame of center of mass when the two objects contact each other, respectively, $\mu$ is the reduced mass $M_\text{imp}M_\text{tar}/M_\text{tot}$, and $v_\text{imp}$ is the impact velocity ($v_\text{imp} = V_\text{imp} - V_\text{tar}$ for negative $V_\text{tar}$). The subscript $R$ in $Q_R$ means reduced mass. Although the classical definition of the specific energy of impact ($Q = 0.5\ M_{imp}\ v_{imp}^2\ /\ M_{tot}$) has been used frequently, we use $Q_R$ expressed in Eq. (1). Note that $Q_R$ is identical to $Q$ when the impactor is much smaller than the target ($M_{imp} << M_{tar}$).

The critical specific impact energy ($Q_{RD}^*$), which is the specific impact energy required to disperse the target in two or more bodies with the largest body having exactly half the total mass (i.e., $M_{tot}/2$) after the collision, is often used to characterize disruptive collisions (Leinhardt and Stewart, 2012). Disruptive collisions ($Q_R \sim Q_{RD}^*$) have been regarded as an important process for planet formation but non-disruptive small-scale collisions ($Q_R << Q_{RD}^*$) have been frequently underestimated or neglected (e.g., Inaba et al., 2003; Wyatt et al., 2008). Here, we call non-disruptive small-scale collisions "erosive collisions," because the mass ejected is much smaller than the total mass. Recent studies (Kobayashi et al., 2010; Kobayashi and Tanaka, 2010) have suggested that erosive collisions are also important to the growth of planets. The reason that these collisions are also important is mainly because erosive collisions are much more frequent than disruptive collisions. Although disruptive collisions ($Q_R \sim Q_{RD}^*$) have been investigated extensively (e.g., Benz and Asphaug, 1999; Leinhardt and Stewart, 2012; Jutzi, 2015; Movshovitz et al., 2016), erosive collisions ($Q_R << Q_{RD}^*$) have not been investigated well. There is also no reliable scaling model for erosive collisions between planetesimals. There are several reasons why erosive collisions have not been investigated well: (1) very high numerical resolution is needed to numerically resolve a small impactor, (2) erosive collisions had been not regarded as an important process for planet formation, and (3) an extensive parameter search for $Q_R$ is needed for erosive collisions (for example, $Q_R/Q_{RD}^* = 0.01 - 1$) compared to disruptive collisions (just around $Q_R/Q_{RD}^* = 1$).

In this study, we carried out impact simulations of gravity-dominated planetesimals, using the smoothed particle hydrodynamics (SPH) method. We systematically investigated the dependence of collision outcomes, namely ejected mass on the numerical resolution, target size, impact velocity and impact angle for a wide range of specific impact energy ($Q_R$) from disruptive collisions ($Q_R \sim Q_{RD}^*$) to erosive ones ($Q_R << Q_{RD}^*$). Our aim in this study was to investigate the relationship between ejected mass and specific impact energy ($Q_R$) for various impact parameters and construct a reliable impact erosion model of collisions between gravity-dominated planetesimals for future work on planet formation.

In Section 2, the methods for the numerical code and the initial conditions for collisions are introduced. Section 3 presents the numerical simulation results and investigates the dependence of collision outcomes on the impact parameters. In Section 4, we construct a new erosion model based on our numerical simulations and compare our results with those of some previous studies.

# 2. Methods
## 2.1. Numerical Code for Collisions

To perform impact simulations of planetesimals, we used the SPH method (e.g., Lucy, 1977; Monaghan, 1992), which is a flexible Lagrangian method of solving hydrodynamic equations and has been used widely for impact simulations in planetary science. The SPH method can easily process large deformations and shock waves. Our numerical code is the same as the code used in Genda et al. (2015a,b). It includes self-gravity but does not include material strength. Here, we briefly summarize the code.

The mutual gravity is calculated using the standard Barnes–Hut tree method (Barnes and Hut, 1986; Hernquist and Katz, 1989) on a multicore CPU. The computational cost, which is proportional to $N\log N$, allowed us to deal with a large number of SPH particles. Additionally, we applied the modified terms in the equations of motion and energy proposed by Price and Monaghan (2007) to conserve the energy more effectively. In all of our simulations, the error of the total energy was within 0.1% during impact simulation.

Von Neumann–Richtmyer-type artificial viscosity (Monaghan 1992) was introduced to capture shock waves. A parameter set of $\alpha = 1.0$ and $\beta = 2.0$ was applied in the artificial viscosity term, because these values are quite appropriate for dealing with the energy partition between kinetic and internal energy during propagation of the shock waves induced by a planetesimal-sized collision (Genda et al., 2015a).

The Tillotson equation of state (EOS) developed by Tillotson (1962), which has been applied widely to date in previous studies including planet- and planetesimal-size collisional simulations (e.g., Benz and Asphaug, 1999; Canup and Asphaug, 2001; Jutzi et al., 2010; Genda et al., 2012; Citron et al., 2015; Hosono et al., 2016; Rosenblatt et al., 2016), was used in our SPH code. The Tillotson EOS contains 10 material parameters, and the pressure is expressed as a function of the density and the specific internal energy, which is convenient for treating fluid dynamics. In this paper, we assumed that the colliding planetesimals are undifferentiated rock, and we used the Tillotson EOS with the parameter sets of basalt referenced in Benz and Asphaug (1999). For planetary sized collisions that involve vaporization of rock, another sophisticated but complicated EOS such as ANEOS (Melosh, 2007) has been often used (e.g., Canup 2004; Ćuk and Stewart, 2012). However, in this paper, we used the Tillotson EOS, because almost all previous studies on planetesimal collisions have used the Tillotson EOS, which allows us to directly compare our results with their results (Benz and Asphaug, 1999; Jutzi et al., 2010; Genda et al., 2012; Movshovitz et al., 2016).

## 2.2. Initial Conditions for Collisions

As shown in Fig. 1, we simulated impacts between two planetesimal-size objects. We considered three sizes of targets with radii of $R_{tar}$ = 30, 100, and 300 km. If a planetesimal with $R_{tar}$ < ~ 100 km is made up with a monolithic rock, the effects of elastic strength of rock would not be negligible (e.g., Jutzi et al. 2010). However, it is expected that a growing planetesimal is not a monolithic rock, but would be a pile of damaged rocks like a rubble pile, because they have grown through a lot of collisions. In this scenario, neglecting gravity as we have done is reasonable.

In order to investigate the dependence of collision outcomes on the specific impact energy ($Q_R$), we changed the size of the impactor like Benz and Asphaug (1999) did, while Leinhardt and Stewart (2012) varied the impact velocity. For an impactor, we considered various sizes with a mass ratio $\gamma$ (= $M_{tar}/M_{imp}$) ranging from 1 to 10000, which corresponds to the size ratio (= $R_{tar}/R_{imp}$) ranging from 1 to 100, where $R_{imp}$ is the radius of the impactor. Two impact angles, $\theta$ = 0° and 45°, where $\theta$ = 0° corresponds to a head-on impact and $\theta$ = 90° corresponds to a grazing impact, were investigated for most of the cases, except for the detailed discussion about impact angle dependency in Section 3.3. Three impact velocities ($v_{imp}$), 2, 3, and 5 km/s, were considered in this paper. This range of impact velocity corresponds to the typical impact velocity between planetesimals and/or a planetesimal and a protoplanet in the late stage of protoplanet formation. This is because a random velocity of planetesimals (~ impact velocity) is excited by the gravitational stirring of neighbor protoplanets, and increases up to the escape velocity of the protoplanets (~ 2 km/s for a lunar-sized protoplanets and ~ 5km/s for a Mars-sized protoplanets). We separated the target and impactor at a distance between their centers of mass of 1.2 ($R_{tar} + R_{imp}$). The values of $\theta$ and $v_{imp}$ were defined at the moment that the surfaces of the two bodies came into contact. The calculation time was 250 sec for almost all cases, except for the cases of $R_{tar}$ = 300 km for 1000 sec and some cases of $v_{imp}$ = 2 km/s for 500 sec. We confirmed that the collision outcomes (i.e., the total ejected mass) reasonably converged within these calculation times by carrying out some impact simulations for 5000 sec.

We assumed that both the target and the impactor are made of basalt. For the initial configurations, the SPH particles were placed in a three-dimensional lattice (face-centered cubic) with a density of 2700 kg/m$^3$ within a sphere of a fixed size. The number of SPH particles for the impactor $n_{imp}$, which corresponds to the numerical resolution, was fixed at 1000 for the standard case. Because we used the same mass for each SPH particle, the number of SPH particles for the target depended on the mass ratio $\gamma$. For example, for the case of $\gamma$ =

10000, $10^7$ SPH particles were used for the target.

In Genda et al. (2015a), the number of SPH particles for the target $n_{tar}$ was fixed, because they considered only disruptive collisions in which the target entirely and largely deforms. Here, we also consider erosive collisions (i.e., much smaller impactor) in which large deformation due to the impact appears near the impact point and its area depends on the impactor's size. Therefore, fixed $n_{imp}$ should be applied for erosive collisions. In Section 3.1, where the simulations with $n_{imp}$ = 100 and 10000 and also $n_{imp}$ = 1000 are shown, we discuss the resolution dependence of collisional outcomes. All initial conditions for the impact simulations and their collision outcomes are summarized in Tables 1–4 in the Appendix.

## 2.3. Analysis of the Ejected Mass

The mass of the largest body $M_{lrg}$ after the collision was calculated from the collision outcome data, and we defined the ejected mass $M_{ej} = M_{tot} - M_{lrg}$. We calculated $M_{lrg}$ using the following three-step procedure, as Genda et al. (2015a) and Benz and Asphaug (1999) did. First, we used a friends-of-friends (FOF) algorithm to identify clumps of SPH particles. We call these clumps FOF groups. Next, we determined whether the particles in an FOF group are gravitationally bound. We also determined whether the particles that do not belong to any FOF group are gravitationally bound to each FOF group. This procedure was performed iteratively until the particle numbers of the FOF groups converged. We refer to the converged FOF groups as singly gravitationally bound (SGB) groups. We also set the lower limit in the number of SPH particles for FOF and SGB groups to be 10 SPH particles. Thus, when there is no FOF and SGB groups detected, we define $M_{ej}/M_{tot} = 1$, which means that there are no identifiable clumps and the target is completely disrupted.

Finally, we iteratively determined whether each SGB group is gravitationally bound. If two SGB groups are gravitationally bound to each other, we regarded them as a single group called a finally gravitationally bound (FGB) group. We defined the mass of the largest FGB group as the mass of the largest body $M_{lrg}$. Using this procedure, the value of $M_{lrg}$ quickly converges after the passage of the shock and rarefaction waves in the target body.

It is noted that there are several ways to identify $M_{lrg}$ at a certain time after the collision. Movshovitz et al. (2016) used the two analysis codes to determine $M_{lrg}$ for disruptive collisions between planetesimals; one is the same algorithm as ours, and the other is the algorithm developed by Jutzi et al. (2010). They showed that the difference in $M_{lrg}$ between two analysis codes is less than 5%.

## 3. Collision Outcomes

In this section, we show the dependence of initial conditions and impact parameters on the mass ($M_{ej}$) ejected by collisions. First, we examined the dependence of $M_{ej}$ on numerical resolution in Section 3.1. We found that $M_{ej}$ depends on numerical resolution, but $M_{ej}$ can be nicely scaled when the specific impact energy ($Q_R$) is normalized by $Q_{RD}*$ that is calculated for each numerical resolution. Next, we examined the effect of the target size and impact velocity on $M_{ej}$ in Section 3.2. Here, we also found that $M_{ej}/M_{tot}$ can be nicely scaled by $Q_R/Q_{RD}*$. In Section 3.3, we examined the dependence of impact angles on $M_{ej}$. Finally, in Section 3.4, we carried out another type of collision simulations, that is, an impactor hits onto a target with a flat surface to simulate cratering. We found that the results for these collisions are converged to the results obtained for the collisions of $Q_R < 0.01\ Q_{RD}*$ in Section 3.1 to 3.3. In the following sections, we present the details.

### 3.1. Dependence on Numerical Resolution

Figure 2 shows the ejected mass ($M_{ej}$) normalized by the total mass ($M_{tot}$) for various $Q_R$. The collisional outcomes of $R_{tar}$ = 100 km and $v_{imp}$ = 3 km/s for three different numerical resolutions ($n_{imp}$ = 100, 1000, and 10000) with $\theta$ = 0° and 45° are plotted. Higher-resolution cases always result in large $M_{ej}$ at the same $Q_R$. The difference in $M_{ej}/M_{tot}$ for a certain $Q_R$ is within a factor of two between $n_{imp}$ = 100 and 10000.

The critical specific impact energy $Q_{RD}*$ can be calculated by linear interpolation of the two data sets of $Q_R$ across $M_{ej}/M_{tot}$ = 0.5, which is defined by

$$Q_{RD}^* = [Q_R]_1 + \frac{[Q_R]_2 - [Q_R]_1}{[M_{ej}/M_{tot}]_2 - [M_{ej}/M_{tot}]_1} \times \left(0.5 - [M_{ej}/M_{tot}]_1\right), \qquad (2)$$

where subscript 1 and 2 correspond to the data whose value of $M_{ej}/M_{tot}$ is the closest to 0.5 but smaller and larger than 0.5, respectively. The linear dependence of $M_{ej}/M_{tot}$ on $Q_R$ near $Q_{RD}*$ has been already reported (e.g., Benz and Asphaug, 1999; Leinhardt and Stewart 2012; Movshovitz et al., 2016). The values of $Q_{RD}*$ are listed in the bottom row in Table 1 for $n_{imp}$ = 1000 and the bottom row in Table 2 for $n_{imp}$ = 100 and 10000. For the same impact angle, the value of $Q_{RD}*$ decreases as $n_{imp}$ increases. For example, $Q_{RD}*$ = 3.06 × 10$^5$, 2.19 × 10$^5$, and 1.39 × 10$^5$ J/kg for 45° impacts in the cases of $n_{imp}$ = 100, 1000, and 10000, respectively. It is noted that the mass ratios of the $Q_{RD}*$s for the different resolutions are not the same. The dependence of $Q_{RD}*$ on the numerical resolution is caused by the different efficiencies of energy transfer from kinetic to internal energy during the propagation of the shock and rarefaction waves through the impactor and target, as was discussed previously in Genda et al. (2015a). Here we found that non-disruptive collisions ($Q_R < Q_{RD}*$) also have a dependence of collisional outcome on numerical resolution.

Figure 3 shows the same results shown in Figure 2, but $Q_R$ is normalized by each calculated value of $Q_{RD}^*$. The value of $M_{ej}/M_{tot}$ is nicely scaled by $Q_R/Q_{RD}^*$ despite the different numerical resolutions. Therefore, once we obtain the converged value or a reasonable value of $Q_{RD}^*$ by performing very high-resolution simulations, we do not have to consider numerical resolutions for erosive collisions. Hereafter, in the following subsections, we discuss only the results for the collisions with $n_{imp} = 1000$.

### 3.2. Dependence on Target Size and Impact Velocity

Figure 4 shows the dependence of the ejected mass on the target sizes ($R_{tar} = 30, 100$, and 300 km). In this figure, $Q_R$ is normalized by $Q_{RD}^*$, as in Fig. 3. The values of $Q_{RD}^*$ are listed in Table 3 for $R_{tar} = 30$ and 300 km. For example, $Q_{RD}^* = 2.21 \times 10^4$, $2.19 \times 10^5$, and $9.29 \times 10^5$ J/kg for 45° impacts in the cases of $R_{tar} = 30, 100$, and 300 km, respectively. The catastrophic disruption threshold is known to depend strongly on target sizes (e.g., Benz and Asphaug, 1999; Leinhardt and Stewart, 2012), because the gravitational potential energy of the target depends on its size. However, Fig. 4 shows that the ejected mass is clearly scaled if the specific impact energy is normalized by each calculated $Q_{RD}^*$.

Figure 5 shows the dependence on impact velocity ($v_{imp} = 2, 3$, and 5 km/s). Although the kinetic energy varies by almost a factor of six between $v_{imp} = 2$ and 5 km/s, the ejected mass is scaled by the specific impact energy (and also the normalized specific impact energy). This suggests that the collisional outcomes in the gravity regime are scaled by the kinetic energy, and not by the momentum, for collisions between gravity-dominated planetesimals in the range of $v_{imp} = 2 - 5$ km/s. The ejected mass can be fitted by the relatively simple function of $Q_R/Q_{RD}^*$. Such a fitting model is presented in Section 4.3.

### 3.3. Dependence on Impact Angle

Figure 6 shows the dependence on impact angle. More oblique impact results in less erosive collisions. The values of $Q_{RD}^*$ are listed in Table 1. Since there is a factor of three or one order of magnitude difference in $M_{ej}/M_{tot}$ for different impact angles in the case of $Q_R/Q_{RD}^* < 10^{-1}$ (see Figure 7), $M_{ej}/M_{tot}$ cannot be nicely scaled by $Q_{RD}^*$ like it can be scaled in the cases of $n_{imp}$, $R_{tar}$, and $v_{imp}$ shown in the previous subsections. This is because the slope $q$ in $M_{ej}/M_{tot} \sim Q_R^q$ depends on the impact angle and ranges from $q = 0.8$–1.3 for the cases of $Q_R \sim Q_{RD}^*$ (also see Fig. 9). On the other hand, for the case of $Q_R < 0.01\, Q_{RD}^*$ the slope $q$ seems to converge to 1 for all $\theta$. This occurs because the impactor is much smaller than the target, and the effect of target's curvature becomes negligible. We discuss the effect of the target's curvature on collision outcomes in the next subsections.

### 3.4. Impact onto Planar Target

Because the target's curvature becomes negligible for the case of $Q_R < 0.01\, Q_{RD}^*$ in our impact setting (i.e., the constant impact velocity), we carried out additional simulations for collisions of an impactor onto a plane target (see Figure 8 for the impact configuration) to simulate cratering (e.g., Fukuzaki et al. 2010) and compared these results with the previously obtained results. Here, we refer to a collision onto plane target as a "sphere-to-plane collision," whereas we call the previous type of collision shown above a "sphere-to-sphere collision." In the case of a sphere-to-plane collision, we ignored the mutual gravity besides material strength. Although we considered an impactor with a radius of 10 km in the numerical simulations, all hydrodynamic equations can be rewritten in a dimensionless form. Therefore, we can obtain $M_{ej}$ for different impactor size from only one simulation for $R_{imp}$ = 10 km.

For the numerical resolution, $n_{imp}$ = 1000 was used. For the planar target, we considered a half-sphere target with a radius of 200 km (= $20 R_{imp}$), where the impactor strikes the flat plane of the target. Equal-mass SPH particles were also used for the impactor and the planar target. We confirmed that this target size with a radius of 20 $R_{imp}$ was sufficient for convergence of $M_{ej}$ by comparison with the result for a larger target with a radius of 30 $R_{imp}$.

The method used to determine $M_{ej}$ for a sphere-to-sphere collision described in Section 2.3 cannot be simply applied to the case of sphere-to-plane collision because the entire body of the target is not included in this simulation. Instead, we estimated $M_{ej}$ by comparing the velocity of each particle above the ground with the escape velocity of a spherical target with $R_{tar}$ = 100 km. We assumed that a particle with a velocity exceeding the escape velocity is not bounded by the target's gravity. The escape velocity of the target with radius $R_{tar}$ and mass $M_{tar}$ is given by

$$v_{esc} = \sqrt{2G \frac{M_{tar}}{R_{tar}}}, \qquad (3)$$

where $G$ is the gravitational constant. For the spherical target with $R_{tar}$ = 100 km, $v_{esc}$ = 123 m/s.

We carried out six numerical simulations for sphere-to-plane collisions in the cases of $\theta$ = 0°, 15°, 30°, 45°, 60°, and 75°. The resultant values of $M_{ej}$ are listed in Table 5. When we normalized $M_{ej}$ by $M_{tot}$ (= 1.13 × 10$^{19}$ kg), which is the mass of the spherical target with $R_{tar}$ = 100 km, $M_{ej}/M_{tot}$ = 1.75 × 10$^{-2}$, 1.71 × 10$^{-2}$, 1.55 × 10$^{-2}$, 1.22 × 10$^{-2}$, 7.13 × 10$^{-3}$, and 2.23 × 10$^{-3}$ for $\theta$ = 0°, 15°, 30°, 45°, 60°, and 75°, respectively. The value of $Q_R$ for this sphere-to-plane impact can be estimated to be $4.49 \times 10^3$ J/kg, assuming $M_{tot} = 1.13 \times 10^{19}$ kg. Figure 9 shows the results for the sphere-to-plane collisions (small dots) as well as the results for the sphere-to-sphere collisions (data points).

As mentioned before, $M_{ej}$ can be scaled by $R_{imp}$ (i.e., $M_{ej} \propto M_{imp} = R_{imp}^3$). Therefore, in the case of $R_{imp} = 1$ km, for example, $M_{ej}/M_{tot} = 1.75 \times 10^{-5}$ for $\theta = 0°$. Because $Q_R = 0.5 M_{imp} v_{imp}^2 / M_{tot}$ for a smaller impactor, the slop $q$ in $M_{ej}/M_{tot} \propto Q_R^q$ should be 1 for constant $v_{imp}$. The results for these scalings are also drawn in Figure 9 as lines. Indeed, the results for sphere-to-sphere collisions for the case of $Q_R < 0.01\ Q_{RD}^*$ are converged to the results for sphere-to-plane collisions.

## 4. Discussion
### 4.1. Effect of Curvature on Impact Angle

For erosive collisions ($Q_R \ll Q_{RD}^*$) with high impact velocities examined in this paper, the collision outcomes for sphere-to-sphere collisions converge to those for sphere-to-plane collisions, because the effect of the curvature becomes negligible. On the other hand, the effect of the target's curvature on the collision outcomes is significant for the collisions with $M_{ej}/M_{tot} > 0.01$ or $Q_R/Q_{RD}^* > 0.01$ (see Fig. 9). For the collisions with $\theta < 45°$, $M_{ej}/M_{tot}$ for sphere-to-sphere collisions tends to be larger that for sphere-to-plane collisions. On the other hand, for the collisions with $\theta > 45°$, $M_{ej}/M_{tot}$ tends to be smaller.

Figure 10 shows the regions in the impactor and target where the SPH particles that eventually escape originally come from. Since the impact velocity was set to be 3 km/s for all simulations shown in Fig. 10, the value of $Q_R$ is the same for the cases with the same value of $\gamma$, while $Q_R$ is different among the cases with $\gamma = 100$, 10000, and infinity. If the volumes of the escape region normalized by the impactor's volume are the same (i.e., $M_{ej}/M_{imp}$ = const.) among collisions with different $\gamma$ (= $M_{tar}/M_{imp}$), $M_{ej}/M_{tot} \propto M_{imp}/M_{tot} \sim M_{imp}/M_{tar}$ for $M_{tar} \gg M_{imp}$. Since $Q_R \propto M_{imp}/M_{tar}$ for a constant impact velocity, the slope $q$ in $M_{ej}/M_{tot} \propto Q_R^q$ should be 1. However, in the cases with $\theta = 0°$ (Fig.10A), the normalized volume of the escape region for $\gamma = 100$ is much larger than those for $\gamma = 10000$ and $\infty$ (a sphere-to-plane collision). Therefore, for near head-on collisions ($\theta \sim 0°$), the effect of the target's curvature greatly enhances the ejections (i.e., $q > 1$). On the other hand, in the cases with $\theta = 60°$ (Fig. 10C), the normalized volume of the escape region for $\gamma = \infty$ is slightly larger than those for $\gamma = 100$ and 10000, because the length of the downrange escape region in the target for $\gamma = \infty$ is longer that those for $\gamma = 100$ and 10000. Therefore, for near grazing collisions ($\theta \sim 90°$), the effect of the target's curvature reduces the ejections (i.e., $q < 1$).

The effect of the target's curvature on the ejected mass is explained schematically in Fig.

11. In the case of near head-on sphere-to-plane collision ($\theta \sim 0°$), the target's material along streamline #1 can eject and escape, while the lengths of streamlines #2 and #3 in the target are so long that the materials cannot exceed the escape velocity when they are ejected above the ground. On the other hand, in the case of near head-on sphere-to-sphere collision, the lengths of streamlines #2 and #3 in the target are short due to the target's curvature, with the result that the target's materials along all streamlines #1–3 can escape. Therefore, the effect of the target's curvature greatly enhances the ejected mass for near head-on collisions ($\theta \sim 0°$).

In the case of near grazing collision ($\theta \sim 90°$), the situation becomes different. In both cases of sphere-to-sphere collision and sphere-to-plane collision, the target's materials along streamline #1 can escape. However, due to the target's curvature, the volume above streamline #1 for a sphere-to-plane collision is much larger than that for a sphere-to-sphere collision. Although the material along streamline #2 for a sphere-to-sphere collision exceeds the escape velocity due to the effect of curvature, which is the same effect as appeared in the near head-on collision, the contribution of this effect to total escaping volume is not very large. As a result, the effect of the target's curvature reduces the ejected mass for near grazing collisions ($\theta \sim 90°$).

In the case of nominal oblique collision ($\theta \sim 45°$), the positive and negative effects of the target's curvature cancel each other. As shown in Fig. 9, this cancellation is obvious because the results for the 45° sphere-to-sphere and sphere-to-plane collisions are very similar in the entire range from erosive collisions to disruptive collisions.

## 4.2. Estimation of Converged Values for $Q_{RD}^*$

Genda et al. (2015a) found that $Q_{RD}^*$ depends on the numerical resolutions, and discussed the reason as follows. During the passage of the shock and rarefaction waves, the impact energy is distributed among the internal energy, the kinetic energy, and the gravitational potential. The evolution of the total kinetic energy depends on the resolution in the SPH method, which means that the efficiency of the energy transfer depends on the resolution. Higher total kinetic energy remains after the impact for the case of higher-resolution simulations. Genda et al. (2015a) also confirmed the following relation:

$$Q_{RD}^* = a + bn^{-1/3}, \qquad (4)$$

where $n$ is the number of SPH particles in the numerical simulations, which is related to the numerical resolution. They applied the number of SPH particles used in the target ($n_{tar}$). The value of the coefficient $a$ should be a converged value of $Q_{RD}^*$, because the second term in Eq. (4) becomes zero in the limit of $n \to \infty$. Here, we define the converged value of $Q_{RD}^*$ as $Q_{RD}^*{}_{inf}$ ($= a$). They confirmed Eq. (4) for only one case of a head-on collision ($\theta = 0°$) with $R_{tar}$

= 100 km and $v_{imp}$ = 3 km/s. Here, we check this relation for oblique collisions, especially the collisions with $\theta$ = 45°, because the collision outcomes for $\theta$ = 45° are similar to angle averaged results, and thus important.

Figure 12 shows the dependence of $Q_{RD}^*$ on $n_{imp}$ for five different cases of impact conditions with $\theta$ = 45°. Here we also change the mass ratio ($\gamma$) to determine the value of $Q_{RD}^*$. All initial conditions and collision outcomes for determining these $Q_{RD}^*$ are listed in Table 6. Except for some cases with low resolutions, $Q_{RD}^*$ for these impact conditions depends linearly on $n_{imp}^{-1/3}$. The fitting parameters $a$ and $b$ in Eq. (4) can be determined by these data points, and these values for the five different cases are also listed in Table 6.

The converged values ($Q_{RD}^*{}_{inf}$) for $R_{tar}$ = 100 km with $v_{imp}$ = 2, 3, 5 km/s are similar values within a 30% difference (9.79 × 10$^4$, 7.48 × 10$^4$, and 7.14 × 10$^4$ J/kg, respectively), although the difference in the energy related to the impact velocity (i.e., $v_{imp}^2$) is about 6 times between $v_{imp}$ = 2 and 5 km/s. As shown in Section 3.2, this result also means that the collisional outcomes in the gravity regime are scaled by the kinetic energy.

### 4.3. Impact-Angle Averaged Results and Fitting Formulation

Here, we discuss the results averaged over the impact angle. According to Shoemaker (1962), the probability distribution for impact angle between $\theta$ and $\theta + d\theta$ is given by

$$P(\theta)d\theta = 2\sin(\theta)\cos(\theta)\,d\theta, \quad 0° \leq \theta \leq 90°. \tag{5}$$

Using this probability distribution, an angle-averaged ejected mass $\overline{M}_{ej}/M_{tot}$ is defined by

$$\frac{\overline{M}_{ej}}{M_{tot}} = \int_{0°}^{90°} 2\frac{M_{ej}(\theta)}{M_{tot}}\sin(\theta)\cos(\theta)\,d\theta. \tag{6}$$

Using the obtained values of $M_{ej}$ for $\theta$ = 0°, 15°, 30°, 45°, 60°, and 75°, we approximate the above equation as follows:

$$\begin{aligned}
\overline{M}_{ej} = &\, M_{ej}(0°)C(0°, 7.5°) + M_{ej}(15°)C(7.5°, 22.5°) \\
&+ M_{ej}(30°)C(22.5°, 37.5°) + M_{ej}(45°)C(37.5°, 52.5°) \\
&+ M_{ej}(60°)C(52.5°, 67.5°) + M_{ej}(75°)C(67.5°, 82.5°) \\
&+ M_{ej}(90°)C(82.5°, 90°),
\end{aligned} \tag{7}$$

where the function $C(a,b)$ is defined by

$$C(a, b) = \int_a^b 2\sin(\theta)\cos(\theta)\,d\theta. \tag{8}$$

The values of $C$ in Eq. (8) can be calculated simply. Although we did not perform the collision with $\theta$ = 90°, it is obvious that $M_{ej}(90°)$ = 0. The angle averaged $\overline{M}_{ej}/M_{tot}$ for sphere-to-sphere collisions and sphere-to-plane collisions is plotted in Figure 13 and listed in Tables 1 and 5.

Using the relationship between $Q_R$ and $\bar{M}_{ej}/M_{tot}$, we can estimate the angle averaged $\bar{Q}_{RD}*$ to be $1.81 \times 10^5$ J/kg, which is close to $Q_{RD}*$ for $\theta = 45°$ ($= 2.19 \times 10^5$ J/kg). This occurs because the contribution of the impact with $\theta \sim 45°$ to $\bar{M}_{ej}/M_{tot}$, that is, $C(37.5°, 52.5°)$, is the largest, and average of $M_{ej}$ (30°) and $M_{ej}$ (60°) is close to $M_{ej}$ (45°). However, there is another way to estimate the angle averaged $\bar{Q}_{RD}*$. Instead of using the relationship between $Q_{RD}$ and $\bar{M}_{ej}/M_{tot}$, $\bar{Q}_{RD}*$ can be estimated directly by using $Q_{RD}$ for each impact angle on the basis of the same concept used in Eq. (6) (e.g., Benz and Asphaug 1999):

$$\bar{Q}_{RD}* = \int_{0°}^{90°} 2Q_{RD}*(\theta)\sin(\theta)\cos(\theta)\,d\theta. \tag{9}$$

Because $Q_{RD}*(90°)$ should be infinity, we cannot in principle integrate the above equation. However, if we ignore the term of $Q_{RD}*(90°)C(82.5°, 90°)$, $\bar{Q}_{RD}*$ is estimated to be $4.13 \times 10^5$ J/kg, which is twice larger than the previously estimated value of $\bar{Q}_{RD}*$ ($1.81 \times 10^5$ J/kg). Hence, we suggest that the $\bar{Q}_{RD}*$ derived from the relationship between $Q_{RD}$ and $\bar{M}_{ej}/M_{tot}$ should be used if the factor of two difference of $\bar{Q}_{RD}*$ is significant for addressing problems.

Figure 13 shows the calculated $\bar{M}_{ej}/M_{tot}$ as a function of $Q_R/\bar{Q}_{RD}*$ for the sphere-to-sphere collisions and the sphere-to-plane collisions. For reference, the data points for $\theta = 45°$ are also plotted. Aside from the similarity of $\bar{Q}_{RD}*$ and $Q_{RD}*(45°)$, we found that $\bar{M}_{ej}$ is close to $M_{ej}$ (45°) for erosive collisions ($Q_R < \bar{Q}_{RD}*$).

Kobayashi and Tanaka (2010) simply assumed the relation of $\bar{M}_{ej}/M_{tot}$ to $Q_R$ as

$$\frac{\bar{M}_{ej}}{M_{tot}} = \frac{\phi}{1+\phi}, \tag{10}$$

where $\phi$ is the normalized specific impact energy $Q_R/\bar{Q}_{RD}*$. This function is determined by the two conditions that $\bar{M}_{ej}$ is one-half of $M_{tot}$ at $\phi = 1$ by definition and that it is an almost linear function at $\phi \ll 1$ based on the scaling law derived by experimental and analytical studies of a crater (e.g., Melosh, 1989). Equation (10) is also drawn in Fig. 13. This formula is consistent with our numerical results, although there is a difference by a factor of 2 for the erosive collisions ($Q_R \ll \bar{Q}_{RD}*$). Indeed, Eq. (10) satisfies the linear dependence between $\bar{M}_{ej}/M_{tot}$ and $\phi$ for the erosive collisions ($Q_R \ll \bar{Q}_{RD}*$) but it does not contain information about the absolute value of $\bar{M}_{ej}/M_{tot}$. Although the formulation given by Kobayashi and Tanaka (2010) is acceptable, we here propose the following new formulation:

$$\frac{\bar{M}_{ej}}{M_{tot}} = 0.44\phi \times \max(0, 1-\phi) + 0.5\phi^{0.3} \times \min(1, \phi). \tag{11}$$

The first term in this equation represents the linear dependence between $\bar{M}_{ej}/M_{tot}$ and $\phi$ for the erosive collisions ($Q_R \ll \bar{Q}_{RD}*$), and the coefficient of 0.44 was determined to fit the results obtained from the collisions between an impactor and planar target. The second term was determined by the average slope of our numerical data points for disruptive collisions ($Q_R \sim \bar{Q}_{RD}*$), and this term should be 0.5 at $\phi = 1$ by definition. The value of 0.3 in the slope is attributed to the angle-averaging operation in Eq. (6). Although the slope for each angle ranges from 0.8 to 1.3 (see Figure 9), which is consistent with the previous studies (e.g., Benz and Asphaug, 1999; Leinhardt and Stewart, 2012; Movshovitz et al., 2016) where the slope at $Q_R \sim Q_{RD}*$ is around 1, the value of angle averaged slope becomes much smaller than 1. This is because, for example, $M_{ej}/M_{tot} = 1$ near $Q_R = 10^5$ [J/kg] for $\theta = 0°$, 15°, and 30°, and thus the angle averaged value of $M_{ej}/M_{tot}$ calculated in Eq. (6) becomes smaller. This fitting formula is also drawn in Figure 13, and it is very consistent with our numerical simulations for the entire range of $\phi$. In Eq. (11), it should be noted that $\bar{M}_{ej}/M_{tot} = 1$ for $\phi > 10$.

### 4.4. Comparison with Previous Works
*4.4.1. Leinhardt and Stewart (2012)*

Although disruptive collisions ($Q_R \sim Q_{RD}*$) have been investigated well, erosive collisions have not been investigated systematically. Leinhardt and Stewart (2012) carried out simulations of disruptive collisions between gravity-dominated ruble-pile bodies using N-body code with finite-size spherical particles including inelastic collisions among particles, but they listed all collision outcomes including erosive collisions ($Q_R < Q_{RD}*$). We compared our results with their results.

Figure 14 shows the comparison between our results and their results. This figure shows our results for the collisions between the target with $R_{tar} = 100$ km and variously sized impactors with $v_{imp} = 3$ km/s, $n_{imp} = 1000$, and $\theta = 0°$ and 45°, and also shows their results for collisions between a target with $R_{tar} = 10$ km and an impactor with four sizes with various $v_{imp}$ and $\theta = 0°$ and 45°. In the case of $\theta = 0°$, our results are similar to their results, despite the different numerical schemes and different target sizes. On the other hand, in the case of $\theta = 45°$, our results seem to be similar to their results for a smaller impactor ($\gamma = 40$) but different from their other results. This is because the similarly sized collisions (especially $\gamma = 1$ and 4) with a low impact velocity (< 1 km/s) in their simulations result in hit-and-run collisions, which correspond to a flat dependence of $Q_R/Q_{RD}*$ on $M_{ej}/M_{tot}$ shown in Fig. 14(B).

Leinhardt and Stewart (2012) also constructed a general model to estimate $Q_{RD}*$ for impact setting. Table 7 listed $Q_{RD}*$ estimated in our study and the value of $\gamma$ at $Q_R = Q_{RD}*$. For these impact setting (i.e., $M_{tar}$, g at $Q_R = Q_{RD}*$, and $\theta$), $Q_{RD}*$ can be estimated by using a Lenhardt and Stewar (2012) model. These estimated values of $Q_{RD}*$ are also listed in Table 7, and the comparison between $Q_{RD}*$ estimated in our study and Leinhardt and Stewart (2012) is shown in Figure 15. We found that the scaling model constructed by Leinhardt and Stewart (2012) agrees with our numerical simulation results within one order of magnitude difference in $Q_{RD}^*$.

*4.4.2 Movshovitz et al. (2016)*

Recently, a new scaling model of disruptive collisions for gravity-dominated bodies was proposed by Movshovitz et al. (2016). They performed many impact simulations for the targets with $R_{tar}$ = 100–1000 km using the SPH method, which is the same method used in this paper. For head-on disruptive collisions (i.e., $Q_R = Q_{RD}^*$), they found the following linear relation between the impact kinetic energy ($K^*$) and the gravitational binding energy of two colliding objects ($U$):

$$K^* = cU, \qquad (12)$$

where $K^*$ and $U$ are defined as

$$K^* = M_{tot} Q_{RD}^*, \qquad (13)$$

$$U = \frac{3GM_{tar}^2}{5R_{tar}} + \frac{3GM_{imp}^2}{5R_{imp}} + \frac{GM_{tar}M_{imp}}{R_{tar} + R_{imp}}. \qquad (14)$$

From their collision outcomes for head-on collisions, they derived $c = c_0 = 5.5 \pm 2.9$.

For oblique impacts, considering the geometric effect, they proposed the following modified impact kinetic energy ($K_\alpha^*$) instead of using $K^*$:

$$K_\alpha^* = \left(\frac{\alpha M_{tar} + M_{imp}}{M_{tar} + M_{imp}}\right) K^*, \qquad (15)$$

where $\alpha$ is the volume fraction of the impactor intersecting the target given by

$$\alpha = \begin{cases} \dfrac{3R_{imp}l^2 - l^3}{4R_{imp}^3}, & l < 2R_{imp}, \\ 1, & l \geq 2R_{imp}, \end{cases} \qquad (16)$$

where $l = (R_{tar} + R_{imp})(1-\sin\theta)$. Because $\alpha$ is always 1 for a head-on collision, the following general relation is established for an arbitrary impact angle:

$$K_\alpha^* = cU. \qquad (17)$$

From their collision outcomes for $\theta$ = 30° and 45°, they found that $c \sim 2\, c_0$ and $3.5\, c_0$, respectively.

Figure 16 shows the relation between $K_\alpha^*$ and $U$ for the cases of 45° collisions obtained by our study, Movshovitz et al. (2016), and Leinhardt and Stewart (2012). We used the data of converged values of $Q_{RD}^*$ estimated in Section 4.2. Despite the very wide range of $U$ ($R_{tar}$ = 10–1000 km) and the different numerical schemes, the linear dependence given by Eq. (12) can be supported. The values of $K_\alpha^*$ in our study are slightly lower than those in Movshovitz et al. (2016), because we used converged $Q_{RD}*$, which leads to slightly smaller values of $K_\alpha^*$.

Based on the scaling model developed by Movshovitz et al. (2016), we can also estimate $Q_{RD}*$ for impact settings that are considered in our paper. Figure 15 shows the comparison between $Q_{RD}*$ estimated in our study and Movshovitz et al. (2016). We found that the scaling model constructed by Movshovitz et al. (2016) agrees with our numerical simulation results within a factor of 3 difference difference in $Q_{RD}*$.

## 5. Conclusions

Although disruptive collisions, in which the impact energy is so large that more than half of the target materials are eroded, have been investigated well, non-disruptive small-scale collisions (erosive collisions) have not been investigated very much despite their frequent occurrence. Here, we systematically carried out impact simulations of gravity-dominated planetesimals, whose specific impact energies ($Q_R$; Eq. (1)) ranged from disruptive collisions ($Q_R \sim Q_{RD}*$) to erosive collisions (down to $Q_R \sim 0.001 Q_{RD}*$), using the smoothed particle hydrodynamics (SPH) method, where $Q_{RD}*$ is the critical specific impact energy for a disruptive collision.

We found that the relation between the ejected mass normalized by the total mass ($M_{ej}/M_{tot}$) and the specific impact energy ($Q_R$) depends on the numerical resolution ($n_{imp}$), the target radius ($R_{tar}$) and the impact velocity ($v_{imp}$) but that it can be nicely scaled by the critical specific impact energy ($Q_{RD}*$) for the parameter ranges investigated in this paper ($R_{tar}$ = 30–300 km, $v_{imp}$ = 2–5 km/s). Although $M_{ej}/M_{tot}$ depends on the impact angle ($\theta$), we confirmed that the collision outcomes for much less erosive collisions ($Q_R < 0.01\ Q_{RD}*$) converge to the results of impact onto a planar target without curvature and that $M_{ej}/M_{tot} \propto Q_R/Q_{RD}*$ holds. For disruptive collisions ($Q_R \sim Q_{RD}*$), the curvature of the target has a significant effect on $M_{ej}/M_{tot}$. For $\theta < 45º$ and $\theta > 45º$, $M_{ej}/M_{tot}$ becomes more and less disruptive compared to the case of impact onto planar target, respectively.

We also examined the angle-averaged value of $M_{ej}/M_{tot}$, and we found that the numerically obtained relations between angle-averaged $M_{ej}/M_{tot}$ and $Q_R/Q_{RD}*$ were very similar to the cases for $\theta = 45º$ impacts. We proposed a new well-fitted erosion model (Eq. (11)) based on our numerical simulations. The model can be applied in studies of the evolution of the asteroid belt (e.g., Bottke et al., 2005), debris disk formation (e.g., Wyatt 2008), and planet formation (e.g., Kobayashi et al., 2010, 2011; Kenyon and Bromley, 2012). First, we can estimate $Q_{RD}*$ for a given impact condition via the scaling model proposed by Movshovitz et al. (2016), and then we can evaluate $M_{ej}/M_{tot}$ using Eq. (11) with the actual value of $Q_R$ for a given impact condition.

Using the statistical method with the impact erosion model given by Eq. (10) and the $Q_D^*$ values estimated by Benz and Asphaug (1999), Kobayashi et al. (2010) calculated the final mass of protoplanets formed in the protoplanetary disk. They found that the final mass of the protoplanets formed was proportional to $(Q_D^*)^{0.87}$. According to Genda et al. (2015a) and Section 4.2 in this study, $Q_D^*$ (also $Q_{RD}^*$) depends on the numerical resolution, and there is a factor of 2–3 difference between the estimated $Q_D^*$ in Benz and Aspahug (1999) and the numerically converged $Q_D^*$. This difference makes the final mass of protoplanets smaller by a factor of 1.8–2.6, which was discussed previously in Genda et al. (2015b). In addition to the difference of $Q_D^*$, there is a factor of 2 difference in $M_{ej}/M_{tot}$ for the erosive collisions ($Q < Q_D^*$) between Eq. (10) used in Kobayashi et al. (2010) and Eq. (11) derived in this study, which means that the frequent erosive collisions are not more erosive than was expected. Therefore, the difference in $M_{ej}/M_{tot}$ for erosive collisions makes the final mass of the formed protoplanets larger. A detailed calculation for protoplanet formation with Eq. (11) and numerically converged values of $Q_{RD}^*$ is required.

## Acknowledgments

We thank anonymous referees for helpful and constructive comments that improved our manuscript. This work was supported by JSPS KAKENHI Grant Nos. 26287101 and 15K13562.

## APPENDIX

Initial impact conditions and collision outcomes for all impact simulations conducted in this paper are listed here. Tables 1–4 list the initial conditions and the collision outcomes for all sphere-to-sphere collisions, and Table 5 lists these for all sphere-to-plane collisions. Table 5 lists the collision outcomes for the additional simulations that were used to determine the converged values of $Q_{RD}^*$ in Section 4.2. For reference, $\gamma = M_{tar}/M_{imp}$, where $M_{tar}$ and $M_{imp}$ are the masses of the target and the impactor, respectively. $R_{tar}$ is the radius of the target, $v_{imp}$ is the impact velocity, and $n_{imp}$ is the number of SPH particles for the impactor. $\theta$ is the impact angle, where $\theta = 0°$ is a head-on collision. $M_{ej}/M_{tot}$ is the ejected mass normalized by the total mass ($= M_{tar} + M_{imp}$). $\bar{M}_{ej}$ is the angle averaged ejected mass estimated by Eq. (6). $Q_R$ is the reduced specific impact energy defined by Eq. (1), and $Q_{RD}^*$ is the critical specific impact energy where $M_{ej}/M_{tot} = 0.5$.

Table 1 Initial conditions and collision outcomes

| | | | $M_{ej}/M_{tot}$ | | | | | | $\bar{M}_{ej}/M_{tot}$ |
|---|---|---|---|---|---|---|---|---|---|
| $\gamma$ | $M_{imp}$ [kg] | $Q_R$ [J/kg] | $\theta = 0°$ | 15° | 30° | 45° | 60° | 75° | average |

$R_{tar} = 100$ km, $M_{tar} = 1.13 \times 10^{19}$ kg, $v_{imp} = 3$ km/s, $n_{imp} = 1000$

| $\gamma$ | $M_{imp}$ [kg] | $Q_R$ [J/kg] | $\theta=0°$ | 15° | 30° | 45° | 60° | 75° | average |
|---|---|---|---|---|---|---|---|---|---|
| 1 | $1.13 \times 10^{19}$ | $1.13 \times 10^{6}$ | 1.00 | 1.00 | 1.00 | 1.00 | $7.69 \times 10^{-1}$ | $5.40 \times 10^{-1}$ | $8.72 \times 10^{-1}$ |
| 3 | $3.77 \times 10^{18}$ | $8.44 \times 10^{5}$ | 1.00 | 1.00 | 1.00 | 1.00 | $4.72 \times 10^{-1}$ | $2.78 \times 10^{-1}$ | $7.71 \times 10^{-1}$ |
| 10 | $1.13 \times 10^{18}$ | $3.72 \times 10^{5}$ | 1.00 | 1.00 | 1.00 | $7.54 \times 10^{-1}$ | $2.13 \times 10^{-1}$ | $1.04 \times 10^{-1}$ | $6.27 \times 10^{-1}$ |
| 20 | $5.66 \times 10^{17}$ | $2.04 \times 10^{5}$ | 1.00 | 1.00 | $9.73 \times 10^{-1}$ | $4.76 \times 10^{-1}$ | $1.22 \times 10^{-1}$ | $5.57 \times 10^{-2}$ | $5.22 \times 10^{-1}$ |
| 30 | $3.77 \times 10^{17}$ | $1.40 \times 10^{5}$ | 1.00 | 1.00 | $8.81 \times 10^{-1}$ | $3.53 \times 10^{-1}$ | $9.08 \times 10^{-2}$ | $3.72 \times 10^{-2}$ | $4.60 \times 10^{-1}$ |
| 70 | $1.62 \times 10^{17}$ | $6.25 \times 10^{4}$ | 1.00 | $8.46 \times 10^{-1}$ | $4.72 \times 10^{-1}$ | $1.70 \times 10^{-1}$ | $4.69 \times 10^{-2}$ | $1.72 \times 10^{-2}$ | $2.89 \times 10^{-1}$ |
| 100 | $1.13 \times 10^{17}$ | $4.41 \times 10^{4}$ | $7.04 \times 10^{-1}$ | $5.54 \times 10^{-1}$ | $3.11 \times 10^{-1}$ | $1.22 \times 10^{-1}$ | $3.50 \times 10^{-2}$ | $1.24 \times 10^{-2}$ | $1.94 \times 10^{-1}$ |
| 200 | $5.66 \times 10^{16}$ | $2.23 \times 10^{4}$ | $2.33 \times 10^{-1}$ | $1.99 \times 10^{-1}$ | $1.31 \times 10^{-1}$ | $6.07 \times 10^{-2}$ | $1.86 \times 10^{-2}$ | $6.22 \times 10^{-3}$ | $7.98 \times 10^{-2}$ |
| 300 | $3.77 \times 10^{16}$ | $1.49 \times 10^{4}$ | $1.27 \times 10^{-1}$ | $1.14 \times 10^{-1}$ | $8.12 \times 10^{-2}$ | $4.23 \times 10^{-2}$ | $1.42 \times 10^{-2}$ | $4.45 \times 10^{-3}$ | $4.98 \times 10^{-2}$ |
| 1000 | $1.13 \times 10^{16}$ | $4.49 \times 10^{3}$ | $2.75 \times 10^{-2}$ | $2.56 \times 10^{-2}$ | $2.05 \times 10^{-2}$ | $1.28 \times 10^{-2}$ | $5.03 \times 10^{-3}$ | $1.47 \times 10^{-3}$ | $1.30 \times 10^{-2}$ |
| 3000 | $3.77 \times 10^{16}$ | $1.50 \times 10^{3}$ | $7.62 \times 10^{-3}$ | $7.26 \times 10^{-3}$ | $6.08 \times 10^{-3}$ | $4.11 \times 10^{-3}$ | $1.80 \times 10^{-3}$ | $5.11 \times 10^{-4}$ | $3.96 \times 10^{-3}$ |
| 10000 | $1.13 \times 10^{16}$ | $4.50 \times 10^{2}$ | $2.06 \times 10^{-3}$ | $1.99 \times 10^{-3}$ | $1.73 \times 10^{-3}$ | $1.25 \times 10^{-3}$ | $6.11 \times 10^{-4}$ | $1.69 \times 10^{-4}$ | $1.16 \times 10^{-3}$ |
| | | $Q_{RD}^*$ [J/kg] | $3.47 \times 10^{4}$ | $4.08 \times 10^{4}$ | $6.78 \times 10^{4}$ | $2.19 \times 10^{5}$ | $8.70 \times 10^{5}$ | $1.08 \times 10^{6}$ | $1.81 \times 10^{5}$ |

Table 2 Initial conditions and collision outcomes for different resolutions

$R_{tar} = 100$ km, $M_{tar} = 1.13 \times 10^{19}$ kg, $v_{imp} = 3$ km/s

| | | | $M_{ej}/M_{tot}$ | | | |
|---|---|---|---|---|---|---|
| | | | $n_{imp} = 100$ | | $n_{imp} = 10000$ | |
| $\gamma$ | $M_{imp}$ [kg] | $Q_R$ [J/kg] | $\theta = 0°$ | 45° | $\theta = 0°$ | 45° |
| 3 | $3.77 \times 10^{18}$ | $8.44 \times 10^{5}$ | 1.00 | 1.00 | --- | --- |
| 10 | $1.13 \times 10^{18}$ | $3.72 \times 10^{5}$ | 1.00 | $5.84 \times 10^{-1}$ | 1.00 | $8.82 \times 10^{-1}$ |
| 20 | $5.66 \times 10^{17}$ | $2.04 \times 10^{5}$ | 1.00 | $3.71 \times 10^{-1}$ | --- | --- |
| 30 | $3.77 \times 10^{17}$ | $1.40 \times 10^{5}$ | 1.00 | $2.69 \times 10^{-1}$ | 1.00 | $5.04 \times 10^{-1}$ |
| 50 | $2.26 \times 10^{17}$ | $8.65 \times 10^{4}$ | $8.95 \times 10^{-1}$ | $1.70 \times 10^{-1}$ | 1.00 | $3.25 \times 10^{-1}$ |
| 100 | $1.13 \times 10^{17}$ | $4.41 \times 10^{4}$ | $3.51 \times 10^{-1}$ | $9.23 \times 10^{-2}$ | $8.85 \times 10^{-1}$ | $1.74 \times 10^{-1}$ |
| 150 | $7.54 \times 10^{16}$ | $2.96 \times 10^{4}$ | --- | --- | $5.90 \times 10^{-1}$ | $1.15 \times 10^{-1}$ |
| 200 | $5.66 \times 10^{16}$ | $2.23 \times 10^{4}$ | --- | --- | $3.42 \times 10^{-1}$ | $8.73 \times 10^{-2}$ |
| 300 | $3.77 \times 10^{16}$ | $1.49 \times 10^{4}$ | $8.03 \times 10^{-2}$ | $3.40 \times 10^{-2}$ | $1.96 \times 10^{-1}$ | $5.68 \times 10^{-2}$ |
| 1000 | $1.13 \times 10^{16}$ | $4.49 \times 10^{3}$ | $1.78 \times 10^{-2}$ | $1.02 \times 10^{-2}$ | $3.85 \times 10^{-2}$ | $1.69 \times 10^{-2}$ |
| 3000 | $3.77 \times 10^{15}$ | $1.50 \times 10^{3}$ | $5.14 \times 10^{-3}$ | $3.23 \times 10^{-3}$ | $1.08 \times 10^{-2}$ | $5.55 \times 10^{-3}$ |
| 10000 | $1.13 \times 10^{15}$ | $4.50 \times 10^{2}$ | $1.34 \times 10^{-3}$ | $9.23 \times 10^{-4}$ | --- | --- |
| | | $Q_{RD}^*$ [J/kg] | $5.59 \times 10^{4}$ | $3.06 \times 10^{5}$ | $2.69 \times 10^{4}$ | $1.39 \times 10^{5}$ |

Table 3 Initial conditions and collision outcomes for different target sizes

$v_{imp} = 3$ km/s, $n_{imp} = 1000$

| | $R_{tar} = 30$ km, $M_{tar} = 3.05 \times 10^{17}$ kg | | | | $R_{tar} = 300$ km, $M_{tar} = 3.05 \times 10^{20}$ kg | | | |
|---|---|---|---|---|---|---|---|---|
| | | | $M_{ej}/M_{tot}$ | | | | $M_{ej}/M_{tot}$ | |
| $\gamma$ | $M_{imp}$ [kg] | $Q_R$ [J/kg] | $\theta = 0°$ | 45° | $M_{imp}$ [kg] | $Q_R$ [J/kg] | $\theta = 0°$ | 45° |
| 2 | --- | --- | --- | --- | $1.53 \times 10^{20}$ | $1.00 \times 10^{6}$ | 1.00 | $5.54 \times 10^{-1}$ |
| 3 | --- | --- | --- | --- | $1.02 \times 10^{20}$ | $8.44 \times 10^{5}$ | $9.70 \times 10^{-1}$ | $4.35 \times 10^{-1}$ |
| 10 | $3.05 \times 10^{16}$ | $3.72 \times 10^{5}$ | 1.00 | 1.00 | $3.05 \times 10^{19}$ | $3.72 \times 10^{5}$ | $8.67 \times 10^{-1}$ | $1.91 \times 10^{-1}$ |
| 20 | --- | --- | --- | --- | $1.53 \times 10^{19}$ | $2.04 \times 10^{5}$ | $4.82 \times 10^{-1}$ | $1.10 \times 10^{-1}$ |
| 30 | $1.02 \times 10^{16}$ | $1.40 \times 10^{5}$ | 1.00 | 1.00 | $1.02 \times 10^{19}$ | $1.40 \times 10^{5}$ | $2.98 \times 10^{-1}$ | $8.00 \times 10^{-2}$ |
| 100 | $3.05 \times 10^{15}$ | $4.41 \times 10^{4}$ | 1.00 | $8.07 \times 10^{-1}$ | $3.05 \times 10^{18}$ | $4.41 \times 10^{4}$ | $5.58 \times 10^{-2}$ | $2.60 \times 10^{-2}$ |
| 200 | $1.53 \times 10^{15}$ | $2.23 \times 10^{4}$ | $9.83 \times 10^{-1}$ | $5.03 \times 10^{-1}$ | --- | --- | --- | --- |
| 300 | $1.02 \times 10^{15}$ | $1.49 \times 10^{4}$ | $8.95 \times 10^{-1}$ | $3.62 \times 10^{-1}$ | $1.02 \times 10^{18}$ | $1.49 \times 10^{4}$ | $1.38 \times 10^{-2}$ | $8.87 \times 10^{-3}$ |
| 600 | $5.09 \times 10^{14}$ | $7.48 \times 10^{3}$ | $6.02 \times 10^{-1}$ | $1.57 \times 10^{-1}$ | --- | --- | --- | --- |
| 1000 | $3.05 \times 10^{14}$ | $4.49 \times 10^{3}$ | $3.11 \times 10^{-1}$ | $8.82 \times 10^{-2}$ | $3.05 \times 10^{17}$ | $4.49 \times 10^{3}$ | $3.33 \times 10^{-3}$ | $2.61 \times 10^{-3}$ |
| 3000 | $1.02 \times 10^{14}$ | $1.50 \times 10^{3}$ | $5.79 \times 10^{-2}$ | $2.56 \times 10^{-2}$ | $1.02 \times 10^{17}$ | $1.50 \times 10^{3}$ | $9.63 \times 10^{-4}$ | $8.40 \times 10^{-4}$ |
| 10000 | $3.05 \times 10^{13}$ | $4.50 \times 10^{2}$ | $1.42 \times 10^{-2}$ | $7.40 \times 10^{-3}$ | --- | --- | --- | --- |
| | | $Q_{RD}^*$ [J/kg] | $6.43 \times 10^{3}$ | $2.21 \times 10^{4}$ | | $Q_{RD}^*$ [J/kg] | $2.12 \times 10^{5}$ | $9.29 \times 10^{5}$ |

Table 4 Initial conditions and collision outcomes for different impact velocities

| | | $R_{tar}$ = 100 km, $M_{tar}$ = 1.13×10$^{19}$ kg, $n_{imp}$ = 1000 | | | | | |
|---|---|---|---|---|---|---|---|
| | | $v_{imp}$ = 2 km/s | | | $v_{imp}$ = 5 km/s | | |
| $\gamma$ | $M_{imp}$ [kg] | $Q_R$ [J/kg] | $M_{ej}/M_{tot}$ | | $Q_R$ [J/kg] | $M_{ej}/M_{tot}$ | |
| | | | $\theta = 0°$ | 45° | | $\theta = 0°$ | 45° |
| 3 | 3.77×10$^{18}$ | 3.75×10$^5$ | 1.00 | 6.78×10$^{-1}$ | --- | --- | --- |
| 5 | 2.26×10$^{18}$ | 2.78×10$^5$ | 1.00 | 5.21×10$^{-1}$ | --- | --- | --- |
| 10 | 1.13×10$^{18}$ | 1.65×10$^5$ | 1.00 | 3.42×10$^{-1}$ | 1.03×10$^6$ | 1.00 | 1.00 |
| 30 | 3.77×10$^{17}$ | 6.24×10$^4$ | 8.23×10$^{-1}$ | 1.58×10$^{-1}$ | 3.90×10$^5$ | 1.00 | 9.31×10$^{-1}$ |
| 50 | 2.26×10$^{17}$ | 3.84×10$^4$ | 4.56×10$^{-1}$ | 1.02×10$^{-1}$ | --- | --- | --- |
| 100 | 1.13×10$^{17}$ | 1.96×10$^4$ | 1.85×10$^{-1}$ | 5.70×10$^{-2}$ | 1.23×10$^5$ | 1.00 | 5.13×10$^{-1}$ |
| 200 | 5.66×10$^{16}$ | --- | --- | --- | 6.19×10$^4$ | 8.78×10$^{-1}$ | 2.47×10$^{-1}$ |
| 300 | 3.77×10$^{16}$ | 6.62×10$^3$ | 4.38×10$^{-2}$ | 1.96×10$^{-2}$ | 4.14×10$^4$ | 6.65×10$^{-1}$ | 1.46×10$^{-1}$ |
| 500 | 2.26×10$^{16}$ | --- | --- | --- | 2.49×10$^4$ | 3.49×10$^{-1}$ | 8.18×10$^{-2}$ |
| 1000 | 1.13×10$^{16}$ | 2.00×10$^3$ | 1.04×10$^{-2}$ | 5.93×10$^{-3}$ | 1.25×10$^4$ | 9.63×10$^{-2}$ | 3.59×10$^{-2}$ |
| 3000 | 3.77×10$^{15}$ | 6.66×10$^2$ | 3.00×10$^{-3}$ | 1.91×10$^{-3}$ | 4.16×10$^3$ | 2.17×10$^{-2}$ | 1.07×10$^{-2}$ |
| 10000 | 1.13×10$^{15}$ | --- | --- | --- | 1.25×10$^3$ | 5.52×10$^{-3}$ | 3.16×10$^{-3}$ |
| | | $Q_{RD}$* [J/kg] | 4.13×10$^4$ | 2.65×10$^5$ | $Q_{RD}$* [J/kg] | 3.28×10$^4$ | 1.19×10$^5$ |

Table 5 Sphere-to-plane collisions

| $R_{imp}$ = 10 km, $v_{imp}$ = 3 km/s | | |
|---|---|---|
| $\theta$ | $M_{ej}$ [kg] | $M_{ej}/M_{tot}$ |
| 0° | 1.98×10$^{17}$ | 1.75×10$^{-2}$ |
| 15° | 1.94×10$^{17}$ | 1.71×10$^{-2}$ |
| 30° | 1.75×10$^{17}$ | 1.55×10$^{-2}$ |
| 45° | 1.38×10$^{17}$ | 1.22×10$^{-2}$ |
| 60° | 8.06×10$^{16}$ | 7.13×10$^{-3}$ |
| 75° | 2.52×10$^{16}$ | 2.23×10$^{-3}$ |
| $\theta$ (average) | 1.24×10$^{17}$ | 1.10×10$^{-2}$ |

Table 6 Resolution dependence of $Q_{RD}$*

| $R_{tar}$ = 100 km, $v_{imp}$ = 3 km/s, $\theta$ = 45° | | | | |
|---|---|---|---|---|
| $n_{imp}$ | $\gamma$ | $Q_R$ [J/kg] | $M_{ej}/M_{tot}$ | $Q_{RD}$* [J/kg] |
| 100 | 10 | 3.72×10$^5$ | 0.584 | 3.06×10$^5$ |
| | 20 | 2.04×10$^5$ | 0.371 | |
| 1000 | 30 | 1.40×10$^5$ | 0.643 | 2.19×10$^5$ |
| | 70 | 6.25×10$^4$ | 0.286 | |
| 3000 | 20 | 2.04×10$^5$ | 0.568 | 1.78×10$^5$ |
| | 30 | 1.40×10$^5$ | 0.404 | |
| 10000 | 30 | 1.40×10$^5$ | 0.504 | 1.39×10$^5$ |
| | 50 | 8.65×10$^4$ | 0.325 | |
| 30000 | 30 | 1.40×10$^5$ | 0.575 | 1.20×10$^5$ |
| | 40 | 1.07×10$^5$ | 0.452 | |
| 100000 | 30 | 1.40×10$^5$ | 0.628 | 1.08×10$^5$ |
| | 50 | 8.65×10$^4$ | 0.417 | |
| $\infty$ | 58.1 | ($b$ = 1.44×10$^6$) | | 7.48×10$^4$ |

| $R_{\text{tar}} = 30$ km, $v_{\text{imp}} = 3$ km/s, $\theta = 45°$ | | | | |
|---|---|---|---|---|
| $n_{\text{imp}}$ | $\gamma$ | $Q_R$ [J/kg] | $M_{\text{ej}}/M_{\text{tot}}$ | $Q_{\text{RD}}$* [J/kg] |
| 100 | 100 | $4.41\times10^4$ | 0.688 | $3.24\times10^4$ |
|  | 150 | $2.96\times10^4$ | 0.454 |  |
| 1000 | 200 | $2.23\times10^4$ | 0.503 | $2.21\times10^4$ |
|  | 300 | $1.49\times10^4$ | 0.362 |  |
| 3000 | 200 | $2.23\times10^4$ | 0.567 | $1.94\times10^4$ |
|  | 250 | $1.79\times10^4$ | 0.465 |  |
| 10000 | 200 | $2.23\times10^4$ | 0.629 | $1.68\times10^4$ |
|  | 300 | $1.49\times10^4$ | 0.457 |  |
| 30000 | 300 | $1.49\times10^4$ | 0.519 | $1.42\times10^4$ |
|  | 320 | $1.40\times10^4$ | 0.495 |  |
| $\infty$ | 365 | ($b = 1.14\times10^5$) | | $1.10\times10^4$ |

| $R_{\text{tar}} = 300$ km, $v_{\text{imp}} = 3$ km/s, $\theta = 45°$ | | | | |
|---|---|---|---|---|
| $n_{\text{imp}}$ | $\gamma$ | $Q_R$ [J/kg] | $M_{\text{ej}}/M_{\text{tot}}$ | $Q_{\text{RD}}$* [J/kg] |
| 100 | 2 | $1.00\times10^6$ | 0.517 | $9.75\times10^5$ |
|  | 3 | $8.44\times10^5$ | 0.413 |  |
| 1000 | 2 | $1.00\times10^6$ | 0.554 | $9.29\times10^5$ |
|  | 3 | $8.44\times10^5$ | 0.435 |  |
| 3000 | 2 | $1.00\times10^6$ | 0.565 | $9.07\times10^5$ |
|  | 3 | $8.44\times10^5$ | 0.456 |  |
| 10000 | 2 | $1.00\times10^6$ | 0.604 | $8.61\times10^5$ |
|  | 3 | $8.44\times10^5$ | 0.487 |  |
| 30000 | 3 | $8.44\times10^5$ | 0.518 | $8.17\times10^5$ |
|  | 4 | $7.20\times10^5$ | 0.436 |  |
| 100000 | 3 | $8.44\times10^5$ | 0.547 | $7.73\times10^5$ |
|  | 4 | $7.20\times10^5$ | 0.464 |  |
| 300000 | 3 | $8.44\times10^5$ | 0.574 | $7.38\times10^5$ |
|  | 4 | $7.20\times10^5$ | 0.487 |  |
| 1000000 | 4 | $7.20\times10^5$ | 0.504 | $7.14\times10^5$ |
|  | 5 | $6.25\times10^5$ | 0.443 |  |
| $\infty$ | 4.23 | ($b = 3.26\times10^6$) | | $6.96\times10^5$ |

| $R_{\text{tar}} = 100$ km, $v_{\text{imp}} = 2$ km/s, $\theta = 45°$ | | | | |
|---|---|---|---|---|
| $n_{\text{imp}}$ | $\gamma$ | $Q_R$ [J/kg] | $M_{\text{ej}}/M_{\text{tot}}$ | $Q_{\text{RD}}$* [J/kg] |
| 100 | 3 | $3.75\times10^5$ | 0.583 | $3.20\times10^5$ |
|  | 5 | $2.78\times10^5$ | 0.437 |  |
| 1000 | 5 | $2.78\times10^5$ | 0.521 | $2.65\times10^5$ |
|  | 10 | $1.65\times10^5$ | 0.342 |  |
| 3000 | 5 | $2.78\times10^5$ | 0.574 | $2.32\times10^5$ |
|  | 8 | $1.98\times10^5$ | 0.443 |  |
| 10000 | 8 | $1.98\times10^5$ | 0.517 | $1.89\times10^5$ |
|  | 10 | $1.65\times10^5$ | 0.451 |  |
| 30000 | 10 | $1.65\times10^5$ | 0.514 | $1.59\times10^5$ |
|  | 12 | $1.42\times10^5$ | 0.460 |  |
| 100000 | 10 | $1.65\times10^5$ | 0.564 | $1.40\times10^5$ |
|  | 15 | $1.17\times10^5$ | 0.442 |  |
| $\infty$ | 18.4 | ($b = 1.94\times10^6$) | | $9.79\times10^4$ |

| $n_\text{imp}$ | $\gamma$ | $Q_R$ [J/kg] | $M_\text{ej}/M_\text{tot}$ | $Q_\text{RD}^*$ [J/kg] |
|---|---|---|---|---|
| 100 | 50 | $2.40\times10^5$ | 0.632 | $1.82\times10^5$ |
|  | 70 | $1.74\times10^5$ | 0.481 |  |
| 1000 | 100 | $1.23\times10^5$ | 0.513 | $1.19\times10^5$ |
|  | 200 | $6.19\times10^4$ | 0.247 |  |
| 3000 | 100 | $1.23\times10^5$ | 0.560 | $1.07\times10^5$ |
|  | 150 | $8.22\times10^4$ | 0.407 |  |
| 10000 | 100 | $1.23\times10^5$ | 0.618 | $9.51\times10^4$ |
|  | 150 | $8.22\times10^4$ | 0.444 |  |
| 30000 | 130 | $9.47\times10^4$ | 0.541 | $8.58\times10^4$ |
|  | 150 | $8.22\times10^4$ | 0.484 |  |
| 100000 | 150 | $8.22\times10^4$ | 0.513 | $7.97\times10^4$ |
|  | 180 | $6.87\times10^4$ | 0.444 |  |
| $\infty$ | 173 | ($b = 5.08\times10^5$) |  | $7.00\times10^4$ |

$R_\text{tar} = 100$ km, $v_\text{imp} = 5$ km/s, $\theta = 45°$

Table 7 Comparison of estimated $Q_\text{RD}^*$

| impact setting | | | this study | | L&S 2012 | M+ 2016 |
|---|---|---|---|---|---|---|
| $R_\text{tar}$ | $v_\text{imp}$ | $\theta$ | $\gamma$ | $Q_\text{RD}^*$ [J/kg] | $Q_\text{RD}^*$ [J/kg] | $Q_\text{RD}^*$ [J/kg] |
| 100 km | 3 km/s | 0° | 127.8 | $3.47\times10^4$ | $1.20\times10^5$ | $2.50\times10^4$ |
|  |  | 45° | 18.54 | $2.19\times10^5$ | $5.96\times10^4$ | $1.55\times10^5$ |
| 30 km | 3 km/s | 0° | 697.8 | $6.43\times10^3$ | $4.53\times10^4$ | $2.24\times10^3$ |
|  |  | 45° | 201.6 | $2.21\times10^4$ | $1.59\times10^4$ | $7.86\times10^3$ |
| 300 km | 3 km/s | 0° | 19.18 | $2.12\times10^5$ | $2.39\times10^5$ | $2.28\times10^5$ |
|  |  | 45° | 2.430 | $9.29\times10^5$ | $3.88\times10^5$ | $1.86\times10^6$ |
| 100 km | 2 km/s | 0° | 46.41 | $4.13\times10^4$ | $5.24\times10^4$ | $2.51\times10^4$ |
|  |  | 45° | 5.366 | $2.65\times10^5$ | $4.60\times10^4$ | $2.02\times10^5$ |
| 100 km | 5 km/s | 0° | 379.3 | $3.28\times10^4$ | $3.00\times10^5$ | $2.49\times10^4$ |
|  |  | 45° | 10.26 | $1.19\times10^5$ | $1.12\times10^5$ | $9.46\times10^4$ |

Note: All values of $Q_\text{RD}^*$ in this study are taken from the results for the case of $n_\text{imp} = 1000$. For the scaling model in L&S 2012 (Leinhardt and Stewart, 2012), we used $c^* = 1.9$ and $\bar{\mu} = 0.36$, which are the best-fit parameters used in their paper. For the scaling model in M+ 2016 (Movshoviz et al., 2016), we used $c = 5.5$ and 19.25 for $\theta = 0°$ and 45°, respectively, which are also the best-fit parameters used in their paper.

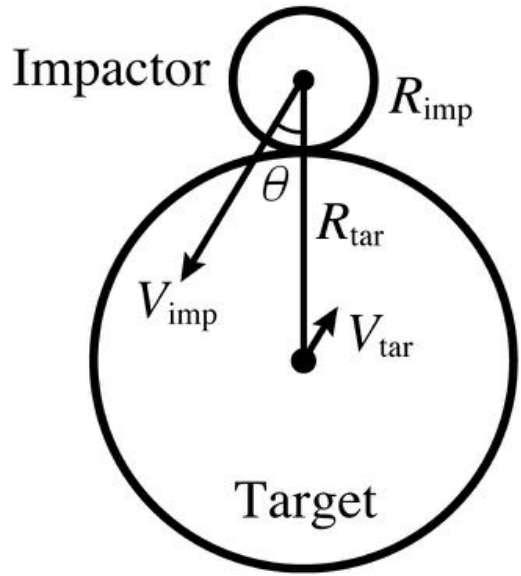

**Figure 1.** Geometries of a collision between a target and an impactor with radii of $R_{tar}$ and $R_{imp}$ ($R_{tar} > R_{imp}$), respectively. The velocities of the target and impactor are $V_{tar}$ and $V_{imp}$, respectively, in the frame of the center of mass. The impact velocity is defined as $v_{imp} = V_{imp} - V_{tar}$ for a negative value of $V_{tar}$. The impact angle is $\theta$, where a head-on collision corresponds to $\theta = 0°$.

- 26 -

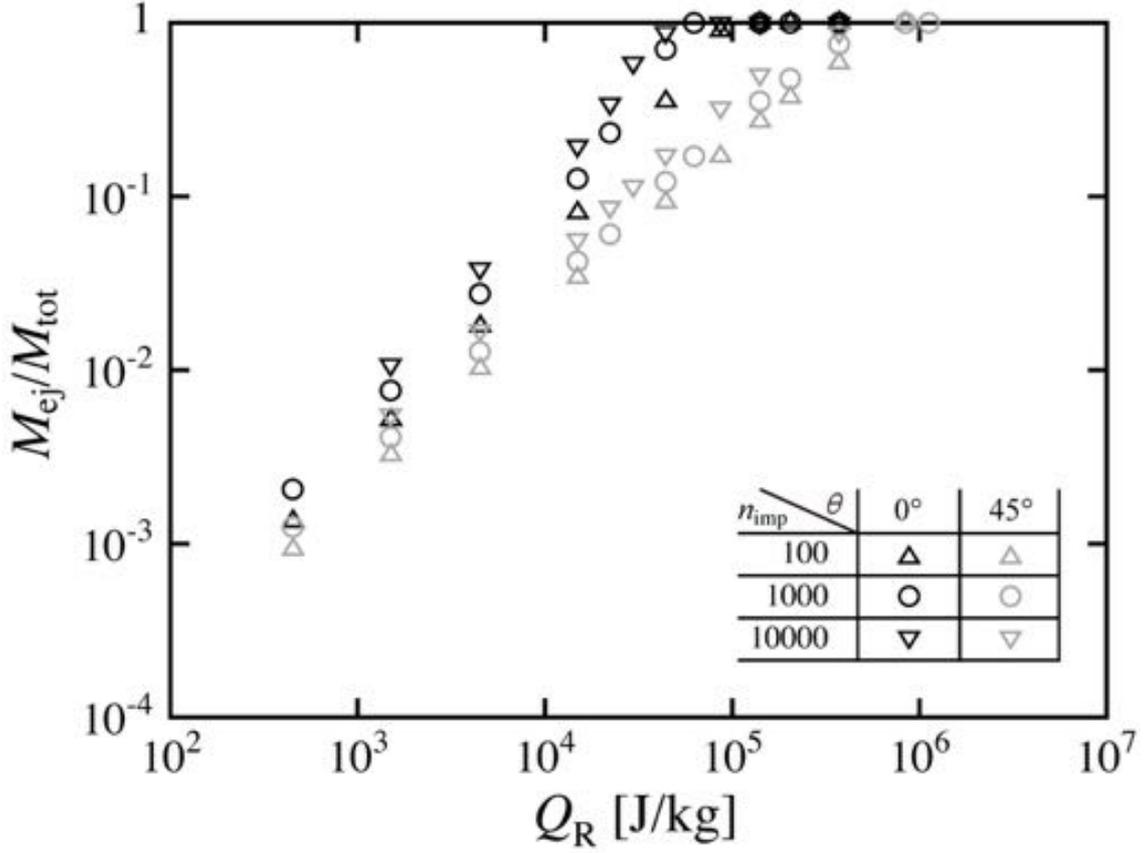

**Figure 2.** Dependence of ejected mass on numerical resolution. The ejected mass normalized by the total mass ($M_{ej}/M_{tot}$) is shown as a function of the specific impact energy ($Q_R$). Black and gray data represent 0° and 45° impacts, respectively, and triangles, circles, and down-pointing triangles represent the calculations with $n_{imp}$ = 100, 1000, and 10000, respectively. The other impact conditions are the same: $R_{tar}$ = 100 km and $v_{imp}$ = 3 km/s. The data listed in Tables 1 and 2 are used.

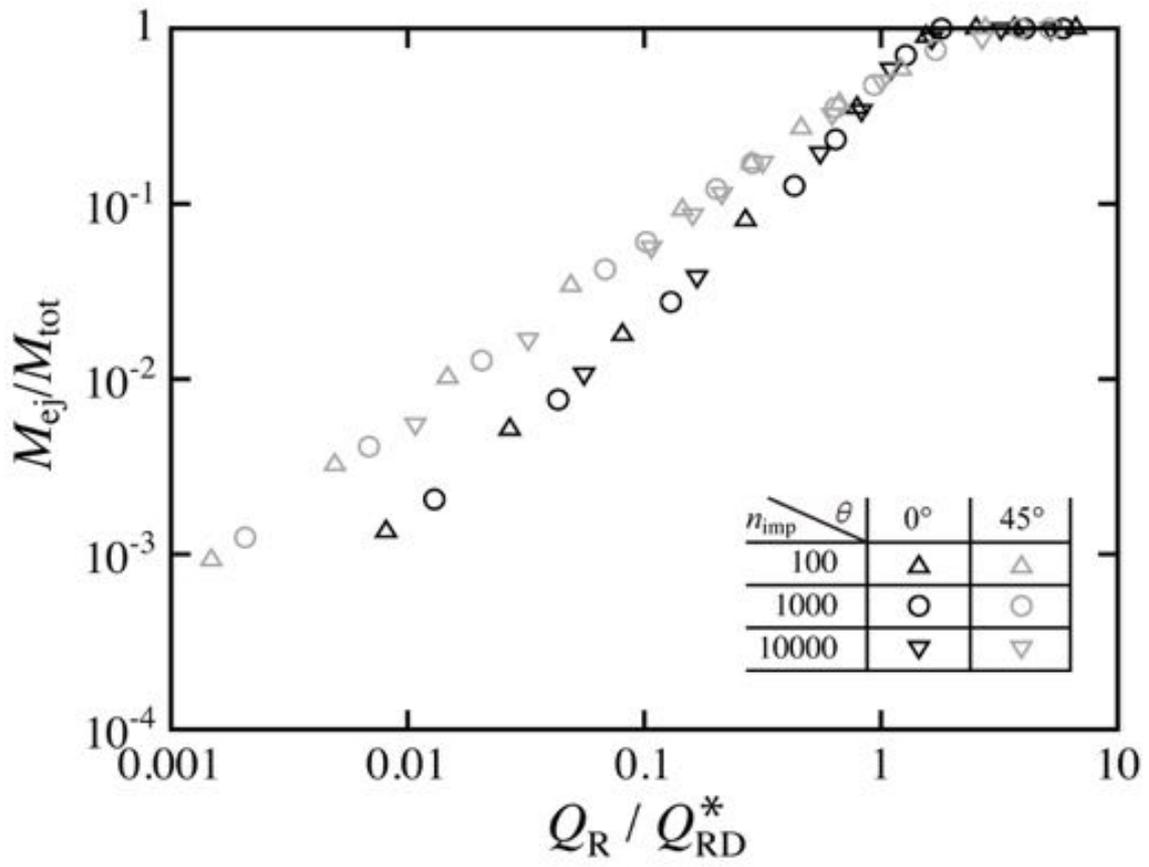

**Figure 3.** Same as Figure 2, but the specific impact energy ($Q_R$) is normalized by $Q_{RD}^*$.

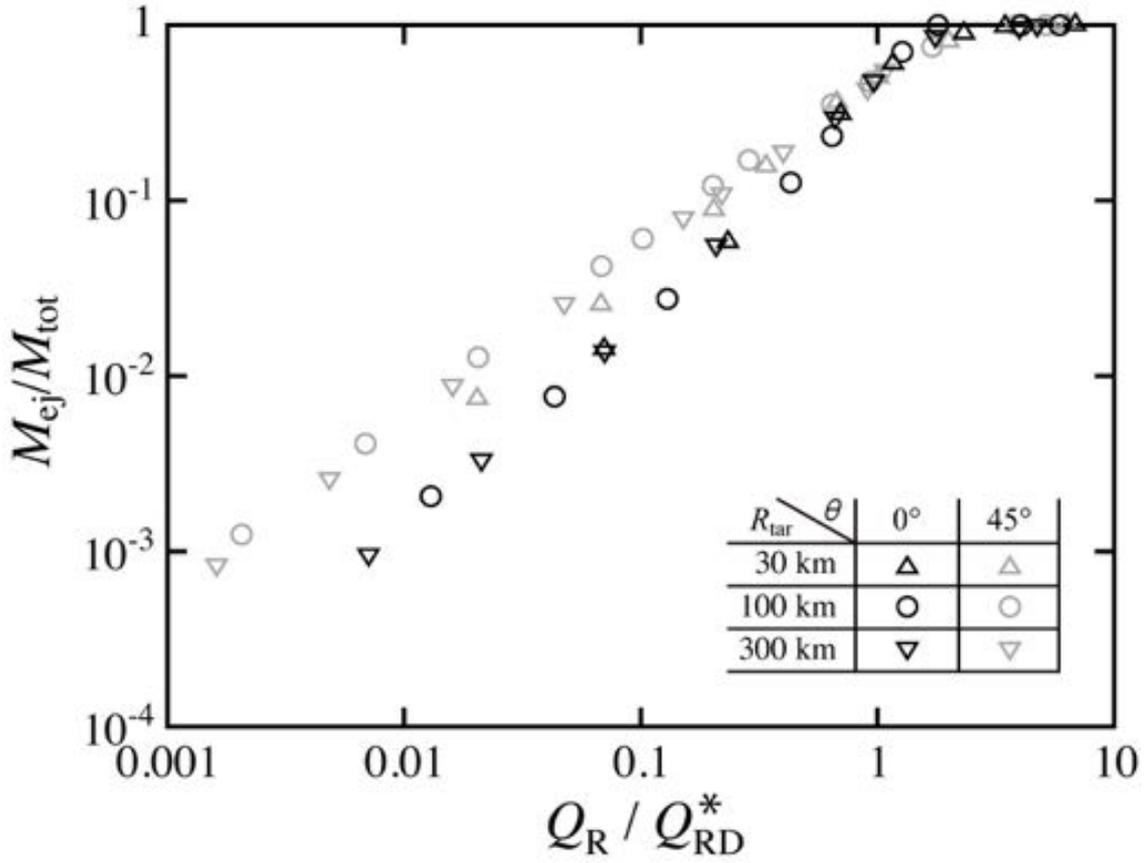

**Figure 4.** Dependence of ejected mass on target size. Black and gray data represent 0° and 45° impacts, respectively, and triangles, circles, and down-pointing triangles represent for target size with $R_{tar}$ = 30, 100, and 300 km, respectively. For all simulations, $v_{imp}$ = 3 km/s and $n_{imp}$ = 1000 are used. The data listed in Tables 1 and 3 are used. The mass ratios ($\gamma$) used in this figure are from 10 to 10000 for $R_{tar}$ = 30 km, from 1 to 10000 for $R_{tar}$ = 100 km, and 2 to 3000 for $R_{tar}$ = 300 km.

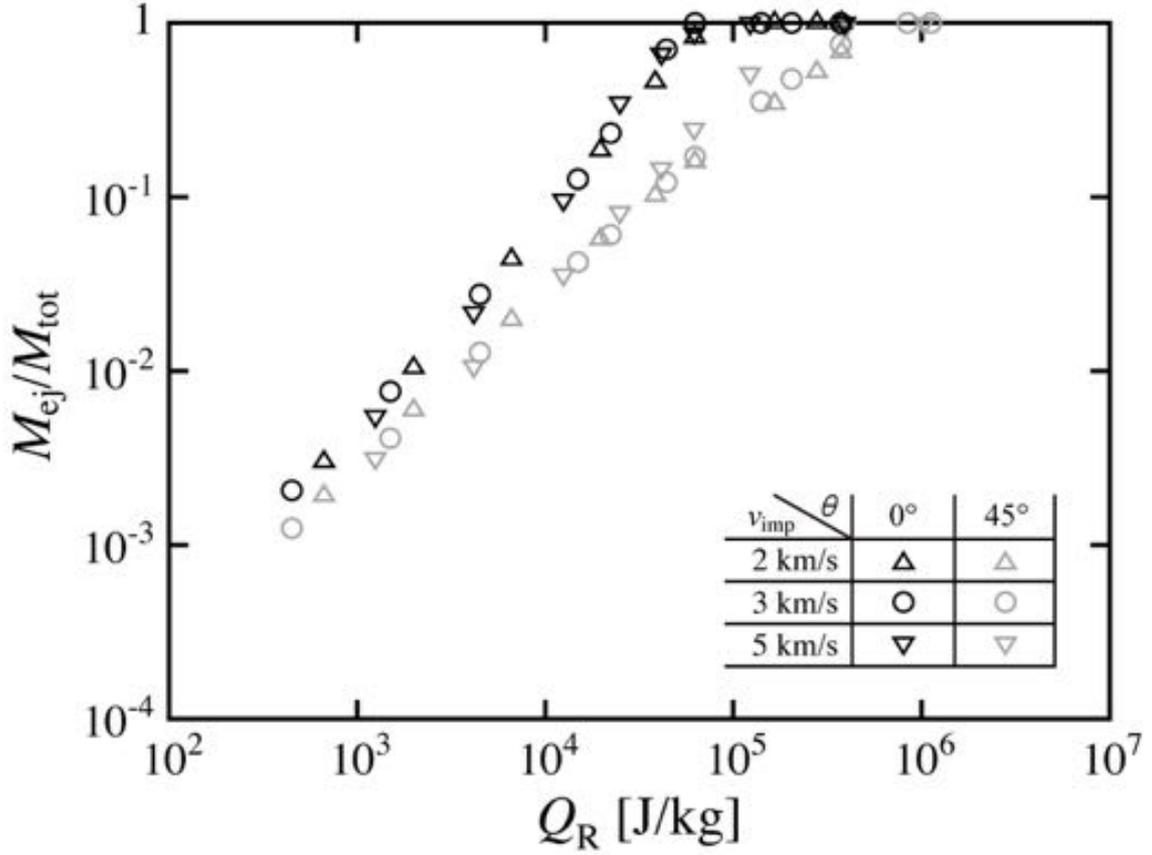

**Figure 5.** Dependence of ejected mass on impact velocity. Black and gray data represent 0° and 45° impacts, respectively, and triangles, circles, and down-pointing triangles represent impact velocity with $v_{imp}$ = 2, 3, and 5 km/s, respectively. For all simulations, $R_{tar}$ = 100 km and $n_{imp}$ = 1000 are used. The data listed in Tables 1 and 4 are used.

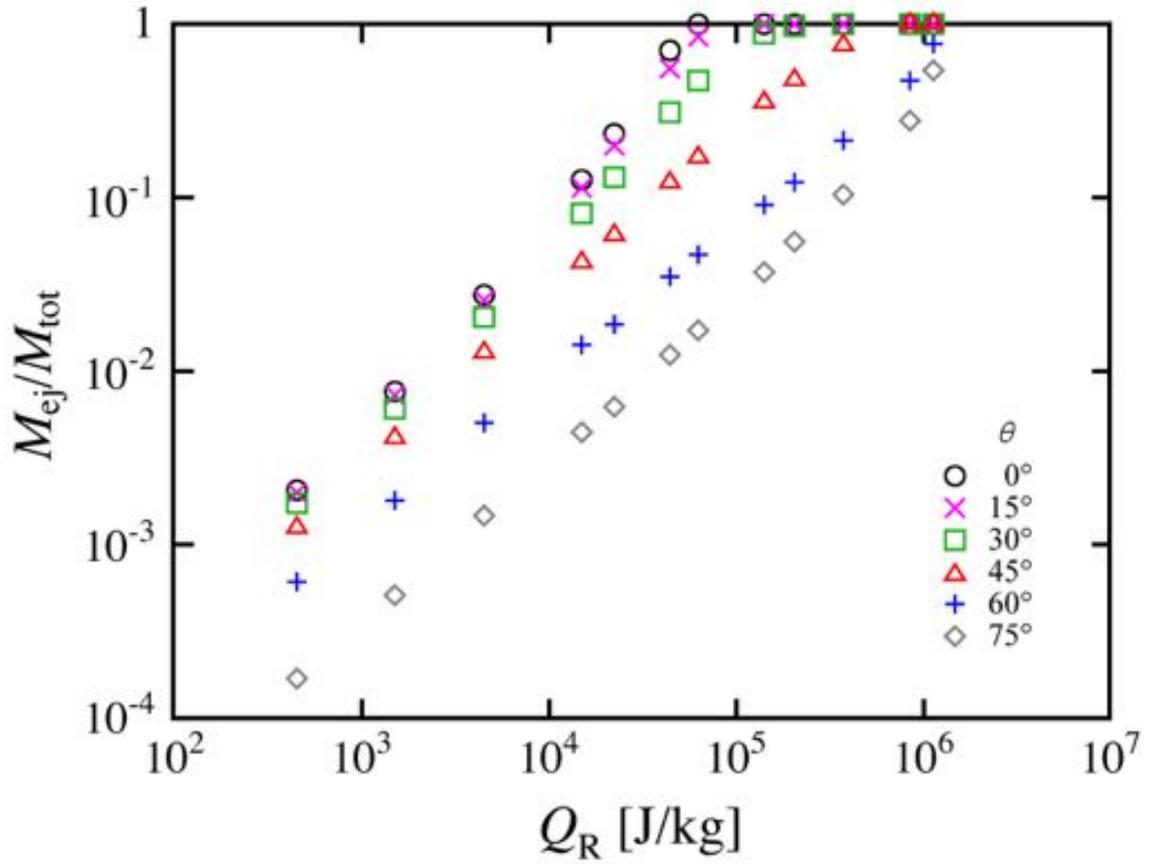

**Figure 6.** Dependence of ejected mass on impact angle. Circles, x marks, squares, triangles, crosses, and diamonds represent for impact angles of $\theta$ = 0°, 15°, 30°, 45°, 60°, and 75°, respectively. For all simulations, $R_{tar}$ = 100 km, $v_{imp}$ = 3 km/s, and $n_{imp}$ = 1000 are used. The data listed in Table 1 are used.

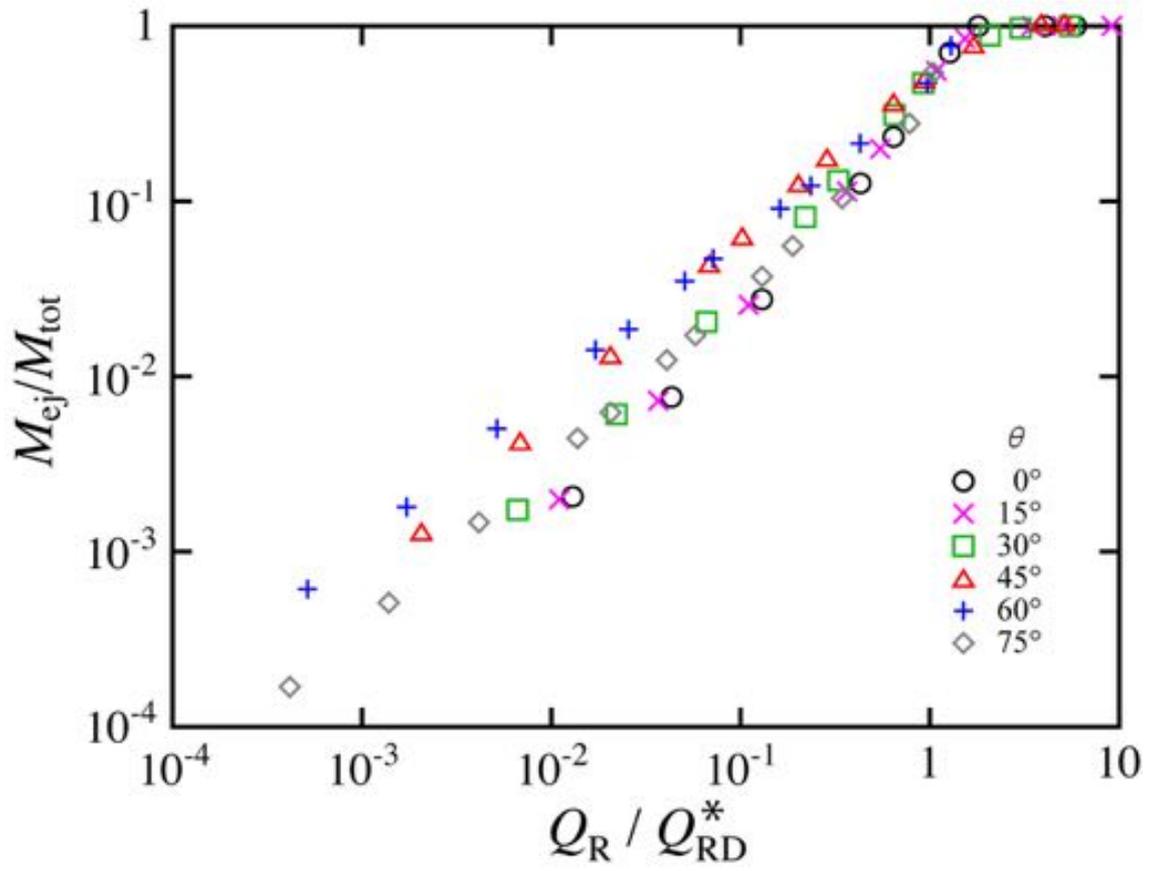

**Figure 7.** Same as Fig. 6, but $Q_R$ is normalized by each $Q_{RD}*$.

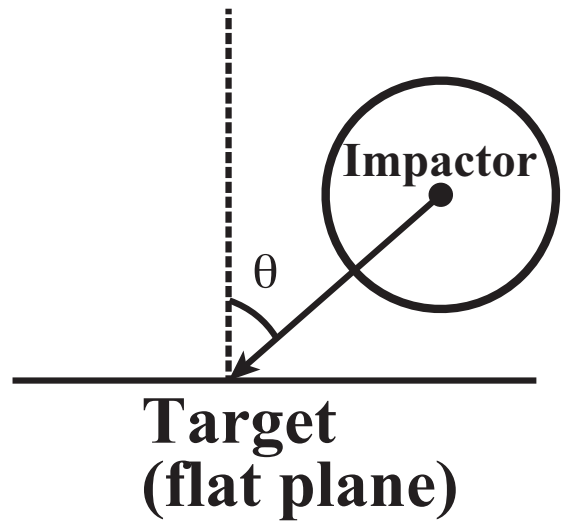

**Figure 8.** Geometries of a sphere-to-plane collision.

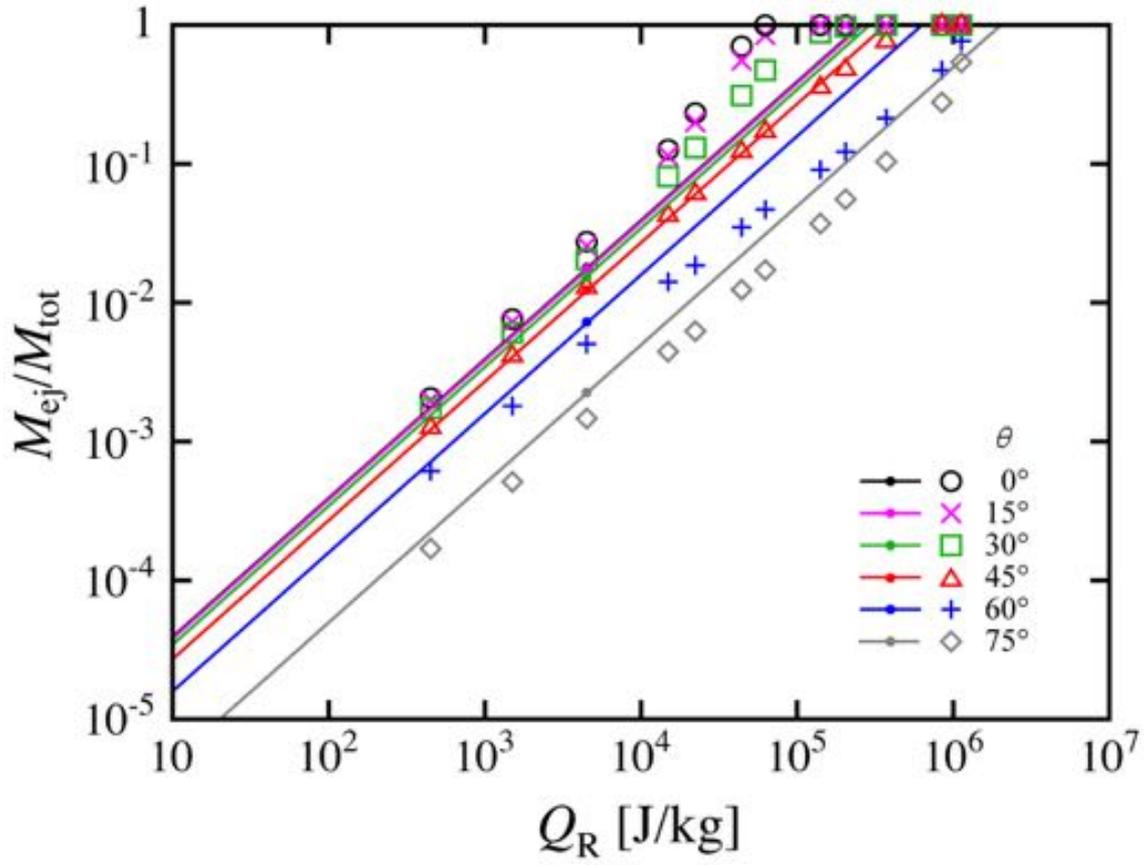

**Figure 9.** Comparison between sphere-to-sphere collisions and sphere-to-plane collisions. Same as Fig. 6, but the results for sphere-to-plane collisions are also shown as the lines. The small dots on these lines are calculated by numerically obtained values of $M_{ej}$ when assuming $R_{tar}$ = 100 km. The data for sphere-to-plane collisions are listed in Table 5.

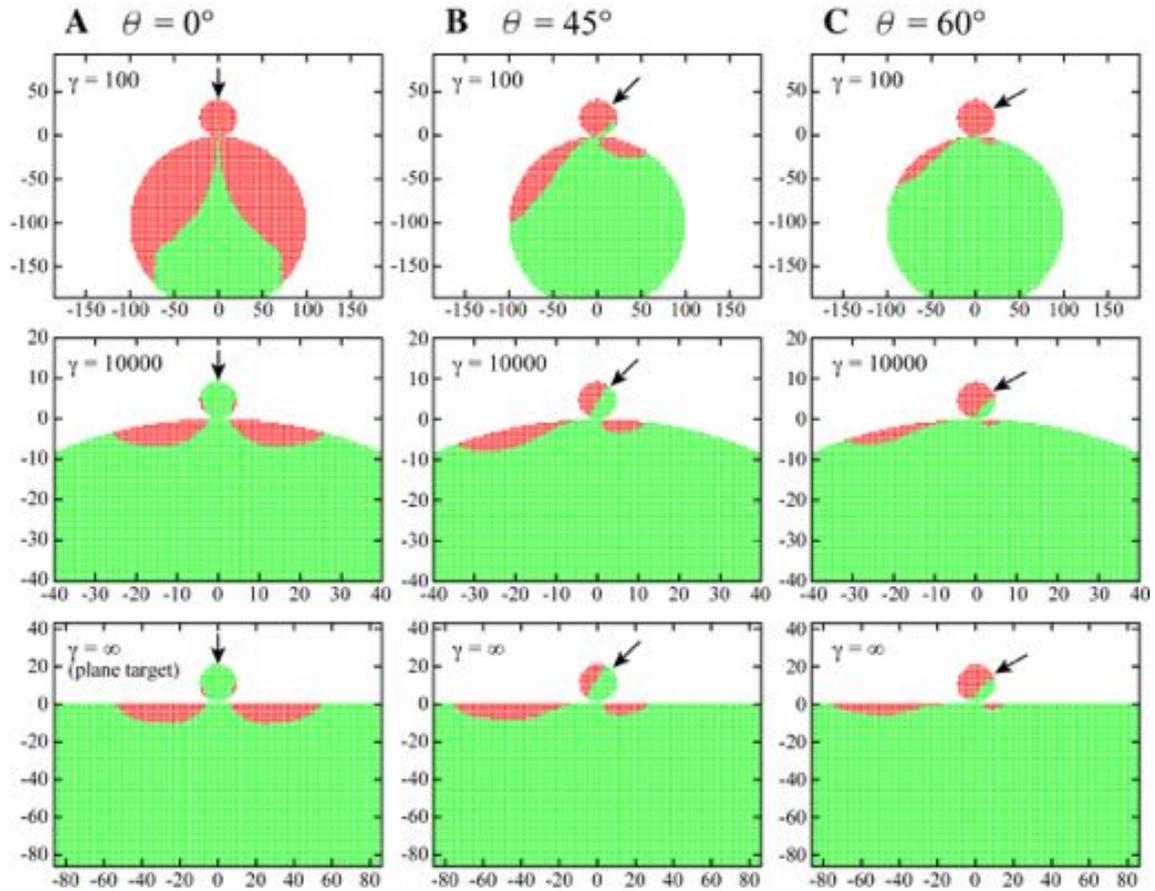

**Figure 10.** Pre-impact locations of escaping SPH particles (red) and bounded SPH particles (green) for the cases of $\theta = 0°$ (A), 45° (B), and 60° (C). Cross-section views on the impact planes are shown. The arrows represent the impact angles. The impactors are drawn at the same size. The bottom panels show the cross sections for sphere-to-plane collisions (i.e., $\gamma = M_{tar}/M_{imp} = \infty$), and the other panels show them for sphere-to-sphere collisions with $\gamma = 100$ and 10000. The unit of the scale is km.

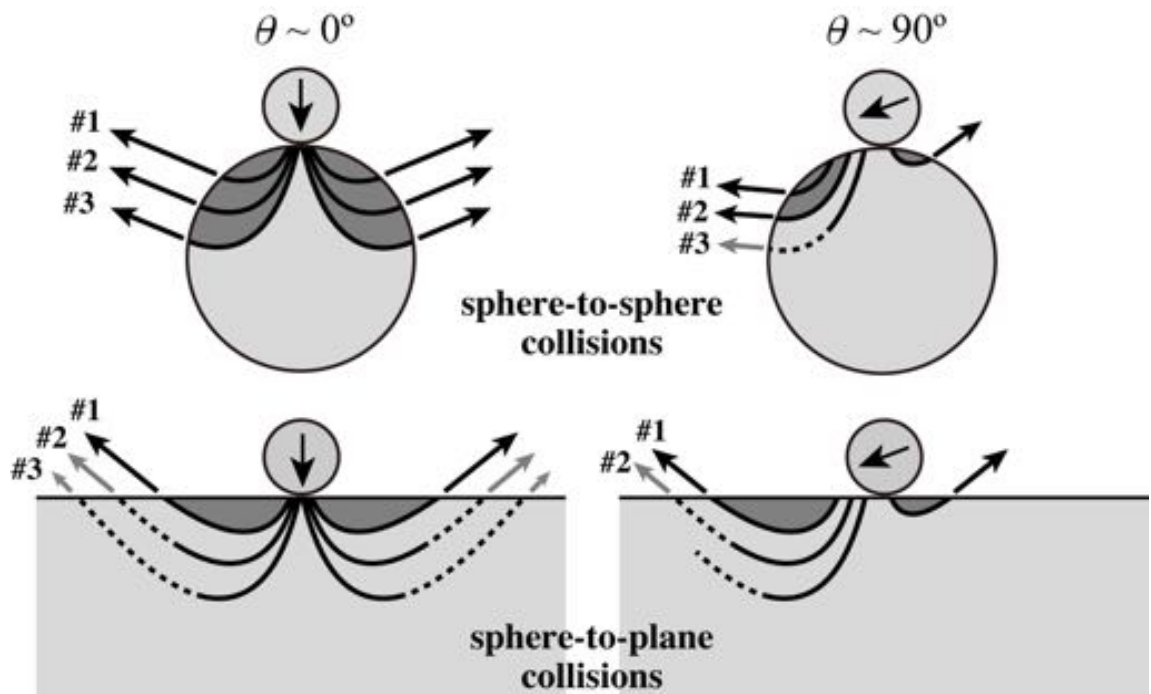

**Figure 11.** Effects of the target curvature on the ejected mass for near head-on collisions ($\theta \sim 0°$) and near grazing collisions ($\theta \sim 90°$). Upper and lower panels represent sphere-to-sphere collisions with a finite target curvature and sphere-to-plane collisions without a target curvature. The curves in the targets represent streamlines of the target materials. The arrows on the targets represent ejection velocities. If the ejection velocity does not exceed the escape velocity, the arrows are colored gray. Dark gray regions in the target represent the area that can escape.

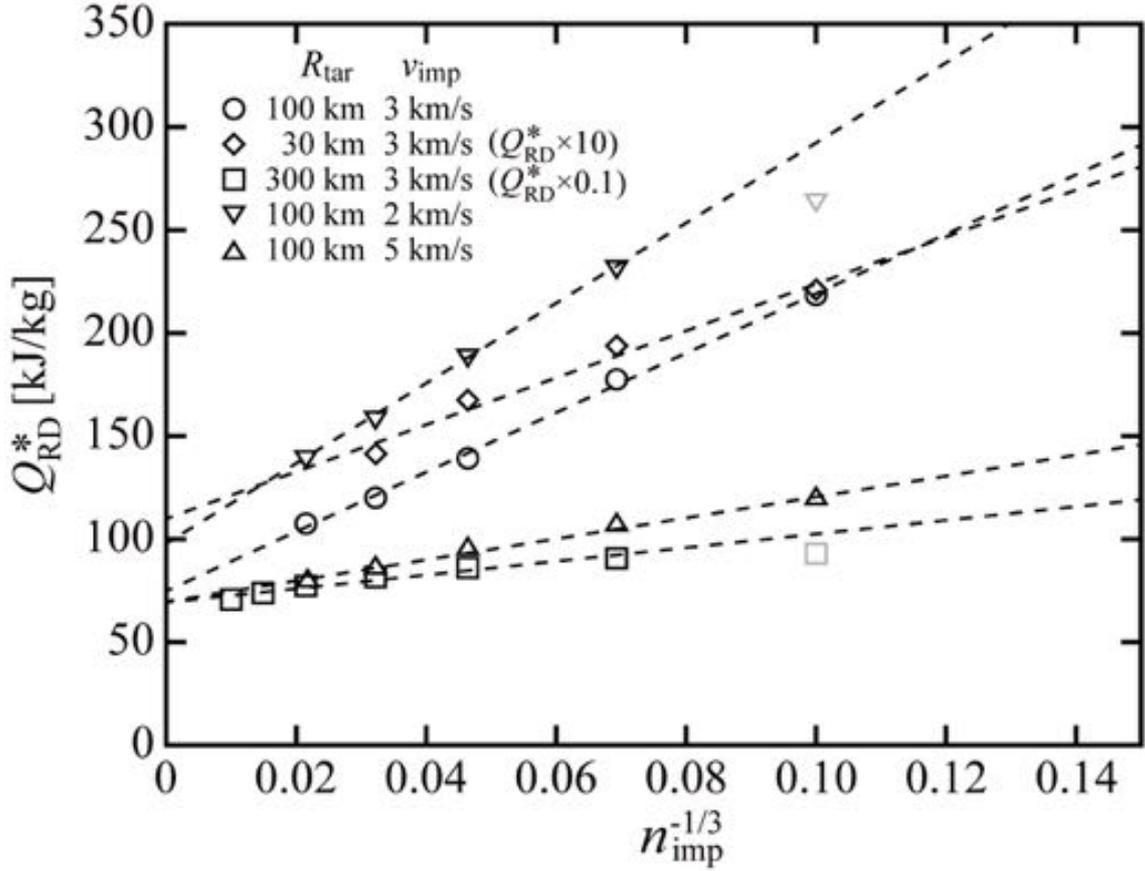

**Figure 12.** The dependence of $Q_{RD}^*$ on numerical resolution ($n_{imp}$) for five different impact conditions with $\theta = 45°$. The dashed curves are the best-fit curves, which are given by Eq. (4). The data for low-resolution simulations (all data for $n_{imp} = 100$ and some data for $n_{imp} = 1000$ that are colored gray) are excluded from these fitting curves. All collision outcomes plotted in this figure are listed in Table 6.

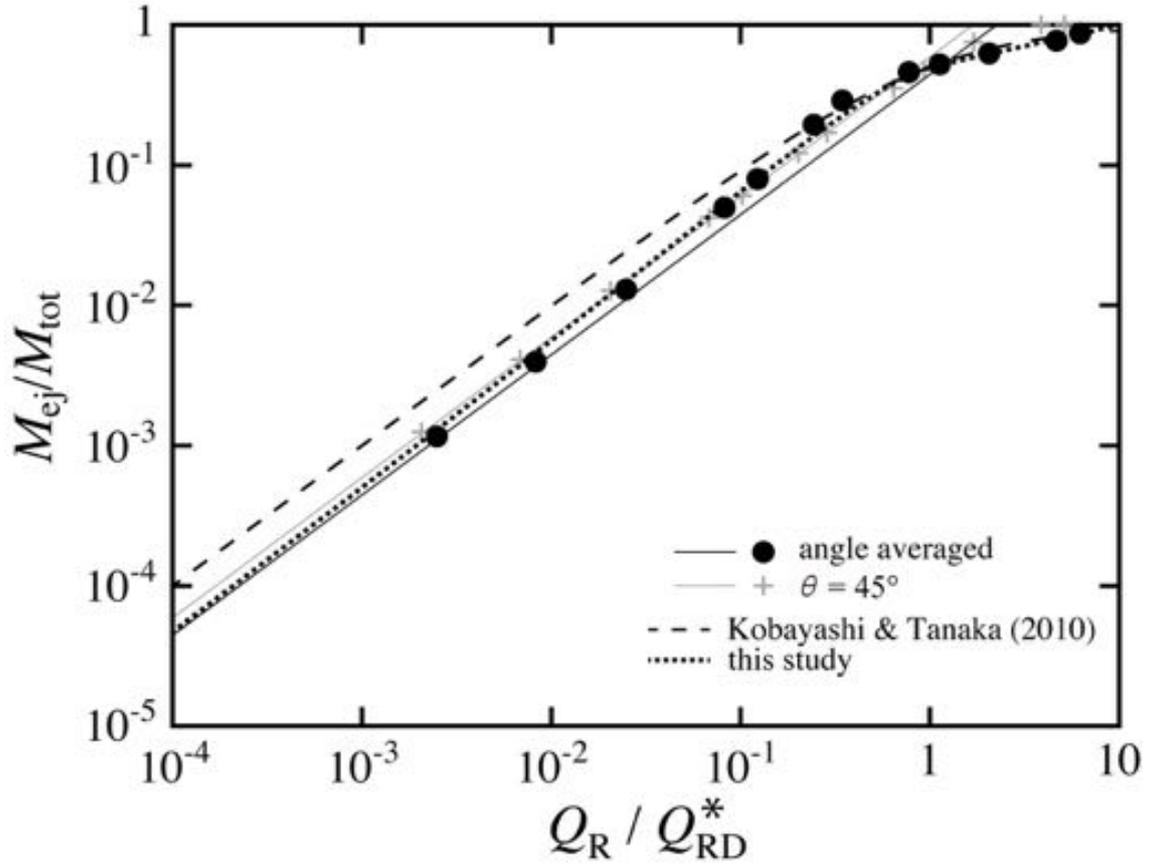

**Figure 13.** The angle averaged ejected mass. Filled circles and solid line are the ejected mass averaged over the impact angles for the sphere-to-sphere collisions and the sphere-to-plane collisions, respectively. Gray crosses and line are the data for $\theta = 45°$. Fitting curves by Kobayashi and Tanaka (2010) (Eq. (10)) and this study (Eq. (11)) are also shown.

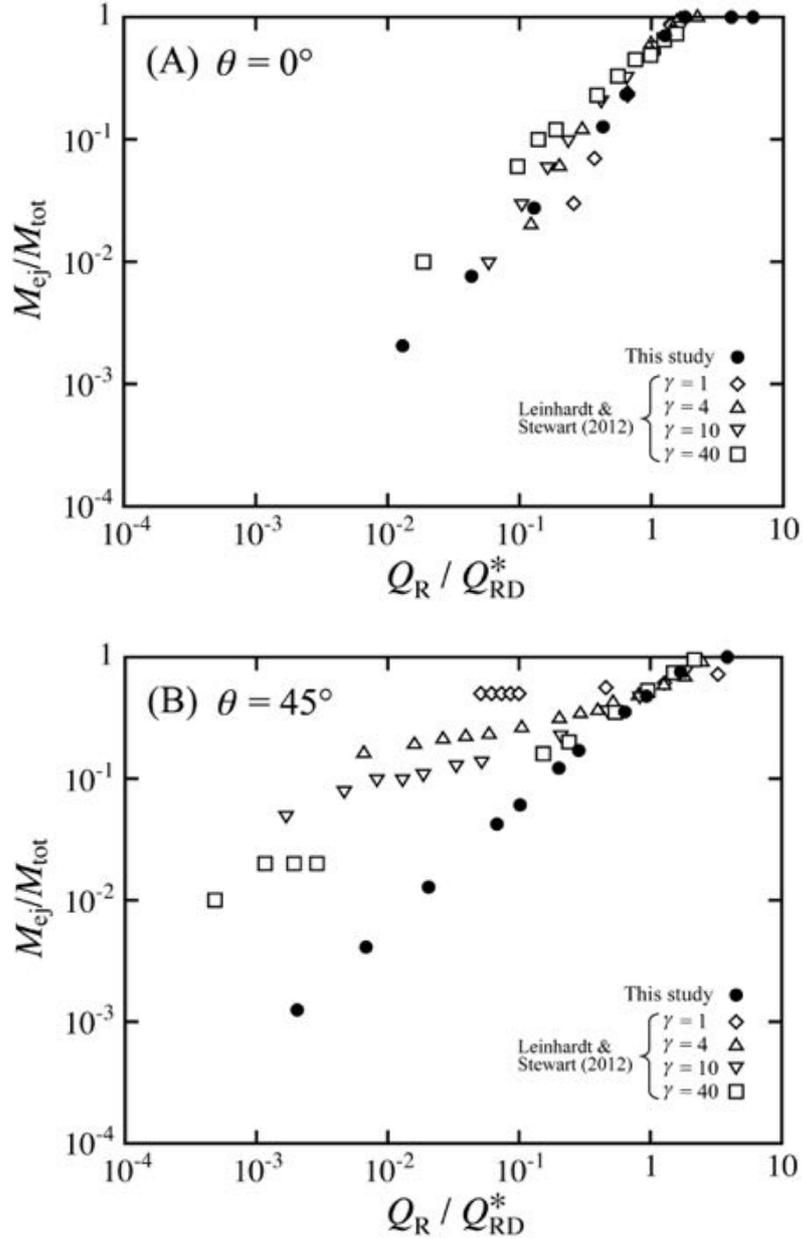

**Figure 14.** Comparison of collision outcomes between Leinhardt and Stewart (2012) and our study for the wide range of $Q_R/Q_{RD}^*$ in head-on collisions (A) and 45-degree collisions (B). Our data are derived from the collisions with $R_{tar} = 100$ km, $v_{imp} = 3$ km/s, and $n_{imp} = 1000$.

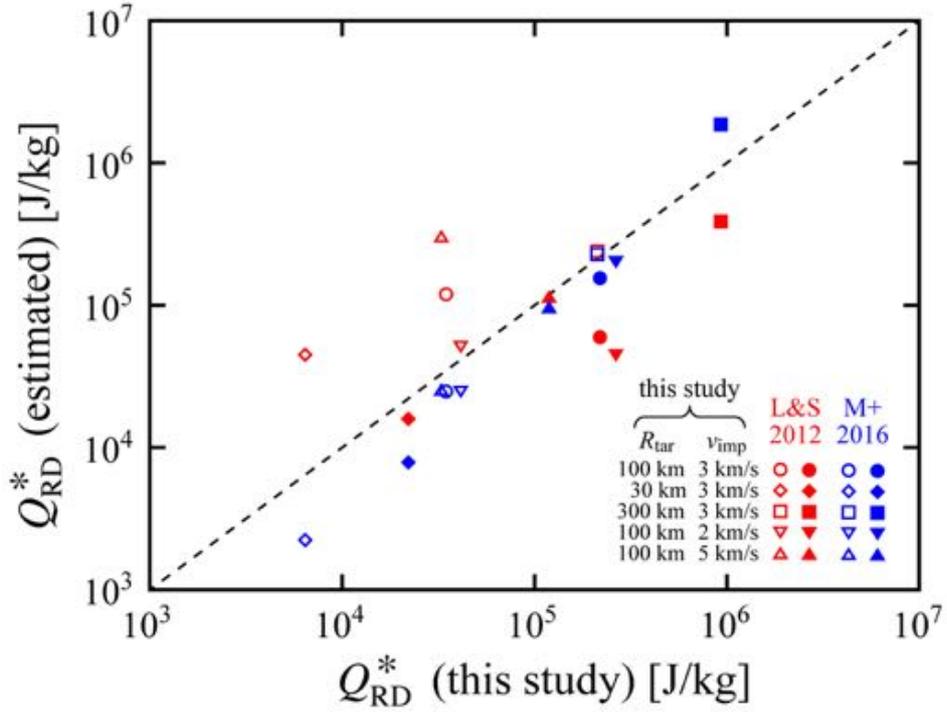

**Figure 15.** Comparison of the values of $Q_{RD}^*$ determined in this study, and estimated by scaling models constructed by Leinhardt and Stewart (2012) and Movshoviz et al. (2016). The values of all data are listed in Table 7. Ten impact settings are considered. Open and filled symbols correspond to the cases of $\theta = 0°$ and $45°$, respectively. Estimated values of $Q_{RD}^*$ by using Leinhardt and Stewart (2012) and Movshoviz et al. (2016) are plotted in red color (L&S 2012) and blue color (M+ 2016), respectively.

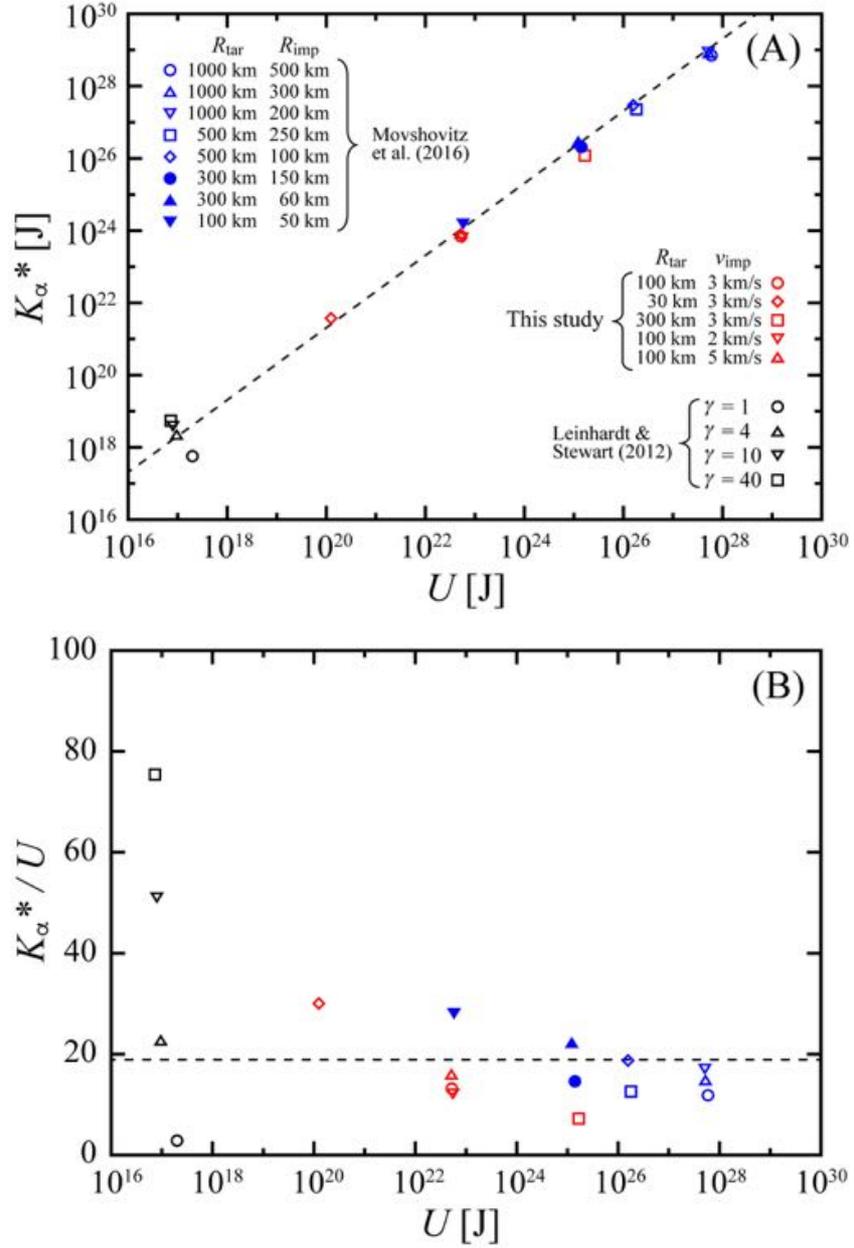

**Figure 16.** The scaling relations between the modified impact kinetic energy ($K_\alpha^*$) and the gravitational binding energy ($U$) in (A) with the values of $K_\alpha^*/U$ plotted instead in (B). Collisions with $\theta = 45°$ performed in this study (red), Movshoviz et al. (2016) (blue), and Leinhardt and Stewart (2012) (black) are plotted. The dashed line is the fitting line ($K_\alpha^* = 19.25\,U$) proposed by Movshoviz et al. (2016).